\documentclass[article,nojss]{jss}
\usepackage{orcidlink,thumbpdf,lmodern}
\usepackage{framed}
\usepackage{amssymb}
\usepackage{amsmath}
\usepackage{hyperref}
\usepackage{dsfont}
\usepackage{enumitem}
\usepackage{graphicx}
\usepackage{float}
\usepackage[section]{placeins}

\DeclareMathOperator*{\argmin}{arg\,min}  % Define \argmin

\author{Maximilian Matthe~\orcidlink{0000-0002-5059-4789}\\Indiana University}
\Plainauthor{Maximilian Matthe}

\title{\proglang{evomap}: A Toolbox for Dynamic Mapping in \proglang{Python}}
\Plaintitle{evomap: A Toolbox for Dynamic Mapping in Python}
\Shorttitle{evomap}

\Abstract{

This paper presents \pkg{evomap}, a \proglang{Python} package for dynamic mapping. 
Mapping methods are widely used across disciplines to visualize relationships among objects as spatial representations, or maps. 
However, most existing statistical software supports only static mapping, which captures objects' relationships at a 
single point in time and lacks tools to analyze how these relationships evolve. 
\pkg{evomap} fills this gap by implementing the dynamic mapping framework EvoMap, originally proposed by 
\cite{Matthe+Ringel+Skiera:2023}, which adapts traditional static mapping methods for dynamic analyses. 
The package supports multiple mapping techniques, including variants of Multidimensional Scaling (MDS), Sammon Mapping, 
and t-distributed Stochastic Neighbor Embedding (t-SNE). It also includes utilities for data preprocessing, exploration, 
and result evaluation, offering a comprehensive toolkit for dynamic mapping applications. 
This paper outlines the foundations of static and dynamic mapping, describes the architecture and functionality of \pkg{evomap}, 
and illustrates its application through an extensive usage example.
}

\Plainkeywords{dynamic mapping, multidimensional scaling, proximity scaling, open-source software, Python}

\Address{
  Maximilian Matthe\\
  Kelley School of Business, Indiana University\\
  Department of Marketing\\
  1309 E 10th St\\ 
  47405 Bloomington, IN, USA\\
  E-mail: \email{mpmatthe@iu.edu}\\
  URL: \url{https://mpmatthe.github.io}
}

\begin{document}

%% -- Introduction -------------------------------------------------------------

%% - In principle "as usual".
%% - But should typically have some discussion of both _software_ and _methods_.
%% - Use \proglang{}, \pkg{}, and \code{} markup throughout the manuscript.
%% - If such markup is in (sub)section titles, a plain text version has to be
%%   added as well.
%% - All software mentioned should be properly \cite-d.
%% - All abbreviations should be introduced.
%% - Unless the expansions of abbreviations are proper names (like "Journal
%%   of Statistical Software" above) they should be in sentence case (like
%%   "generalized linear models" below).

\section[Introduction]{Introduction} \label{sec:intro}

\textit{Mapping}, also referred to as proximity scaling, encompasses a range of statistical methods that analyze data 
consisting of pairwise relationships through spatial representations, often called \textit{maps}. 
These methods project pairwise relationships onto configurations where objects are positioned 
as points within a lower-dimensional space. The resulting maps facilitate exploring the data structure, reveal hidden patterns, 
and help identify key dimensions of difference.

Prominent techniques such as Multidimensional Scaling \citep[MDS;][]{Carroll+Arabie:1998}, 
Sammon Mapping \citep{Sammon:1969}, and t-distributed Stochastic Neighbor Embedding 
\citep[t-SNE;][]{van-der-Maaten+Hinton:2008} are extensively used across disciplines including 
statistics \citep{Saeed+etal:2019}, marketing \citep{DeSarbo+Manrai+Manrai:1994}, 
psychology \citep{Goodwill+Alasdair+Meloy:2019}, 
psychometrics \citep{Hebart+Zheng+Pereira+Baker:2020}, 
ecology \citep{Kenkel+Orloci:1986}, 
network visualization \citep{Mane+Boerner:2004}, 
and political science \citep{Jacoby+Armstrong:2014}, where they help understand 
market competition, social network structures, ecological interactions, cognitive processes, 
latent psychological traits, and political landscapes.

Mapping methods traditionally generate static maps based on data that capture pairwise relationships among objects 
at a specific point in time. However, these relationships often change over time due to evolving consumer perceptions, 
social interactions, or political stances. When longitudinal data on such evolving relationships is available, 
analysts can gain deeper insights into these changes by studying them in dynamic maps. Although dynamic mapping introduces
additional methodological challenges, EvoMap, as recently proposed by \cite{Matthe+Ringel+Skiera:2023}, addresses them
and provides a flexible framework for adapting static mapping methods, like MDS, Sammon Mapping or t-SNE, to accommodate 
longitudinal data.

Thus far, however, statistical software development has focused almost entirely on static mapping. 
In \proglang{Python}, the \pkg{scikit-learn} package offers implementations of static mapping methods 
such as MDS or t-SNE \citep{Pedregosa+etal:2011}. Individual contributions exist that implement specific 
methods, such as \pkg{sammon-mapping} \citep{Perera:2023}, or \pkg{umap-learn} \citep{McInnes+Healy+Melville:2018}. 
In \proglang{R}, the \pkg{stats} package provides a function for Classical Scaling, and the \pkg{MASS} package
provides functions for Sammon Mapping and non-metric MDS \citep{Venables+Ripley:2002}. The
most comprehensive \proglang{R} package for MDS is \pkg{smacof} \citep{DeLeeuw+Mair:2009, Mair+Groenen+DeLeeuw:2022},
which implements various MDS variants, while MDS extensions can be found in \pkg{smacofx} \citep{Rusch+DeLeeuw+Chen+Mair:2024} or 
\pkg{cops} \citep{Rusch+Mair+Hornik:2021,Rusch+Mair+Hornik:2024}. Other static mapping methods can be
found in \pkg{Rtsne} \citep{Krijthe:2015}, which implements t-SNE, or \pkg{vegan} \citep{Oksanen+etal:2025},
 \pkg{ecodist} \citep{Goslee+Urban:2007}, or \pkg{SensoMineR} \citep{Le+Husson:2008}, which implement variants of 
 non-metric MDS.

 These existing packages provide versatile toolsets for static mapping, but their functionality for dynamic mapping is 
 limited. Their functions typically process only a single relationship matrix at a time. When analyzing a temporal sequence of matrices, such independent mapping often creates incomparable solutions that are 
challenging to compare over time \citep{Matthe+Ringel+Skiera:2023}. 
Although the \pkg{smacof} package supports three-way MDS, which allows to process multiple relationship matrices 
simultaneously, it is designed to analyze differences across individuals rather than changes over time. 
Thus, there is a clear gap in software tools dedicated to creating, exploring and evaluating dynamic maps. 

This paper presents the \proglang{Python} package \pkg{evomap}, designed to fill this gap. Developed 
as a software companion to \cite{Matthe+Ringel+Skiera:2023}, the \pkg{evomap} package provides 
a comprehensive toolbox for dynamic mapping. It offers a flexible implementation of the 
homonymous EvoMap framework compatible with various extant static mapping methods. 
The package has a highly modular design, which facilitates the addition of new mapping methods. Further, it adopts the
popular \pkg{scikit-learn} syntax, which ensures ease of use, and includes comprehensive modules for evaluating and 
exploring dynamic mapping results.

This article first provides a brief overview of the methodological background behind static and dynamic mapping 
required to understand \pkg{evomap}'s functionalities. It then details the package's design and features, 
demonstrates its application through a detailed usage example, and explores advanced considerations for practical use. 
The paper concludes with a discussion of potential directions for future development.

\section{Background on mapping} \label{sec:background}

\subsection{Static mapping} \label{sec:static-mapping}

Static mapping positions a set of $n \in \mathbb{N}^+$ objects in a lower-dimensional space such that the distances among
 them closely reflect their relationships. Typically, the input to a static mapping method consists of a single $n \times n$ matrix $D \in \mathbb{R}^{n \times n}$ 
 of pairwise relationships among $n$ objects, typically measured as non-negative pairwise (dis-)similarities\footnote{Although input data can vary, this discussion focuses on dissimilarities, the most prevalent type, for simplicity.}. 
 We generally assume that $D$ is symmetric and contains no missing values, although extensions exist to handle such data.

Commonly, these methods seek a configuration  
\(X = [x_1, \dots, x_n]^\top \in \mathbb{R}^{n \times d}\) of \(n\) objects such that their interpoint distances in 
the \(d\)-dimensional space approximate the given matrix of observed dissimilarities  
\(\delta_{ij}\) in $D$. This configuration serves as the basis for visualization, often in the form of a two-dimensional map.

To enhance interpretability, the resultant map can be augmented with additional features of the analyzed objects. 
For instance, property fitting \citep{DeSarbo+Hoffman:1987} projects external object attributes onto the configuration by 
regressing each attribute onto the map's dimensions. The resulting coefficients, often visualized as 
vectors in the lower-dimensional space, indicate the strength and direction of association between 
each attribute and the spatial layout. As an alternative, one can bind such attributes to visual 
elements like the size or color of each point \citep{Ringel+Skiera:2016}.

Mapping originated in Psychometrics with Torgerson's Classical Scaling \citep{Torgerson:1958}. Since then, research from 
different fields has developed several variants. For instance, \cite{Shepard:1962a, Shepard:1962b} introduced monotonic 
transformations of the input dissimilarities, leading to non-metric MDS \citep{Kruskal:1964a, Kruskal:1964b}. Other 
extensions process measurements from multiple subjects \citep{Carroll+Chang:1970} or link the input dissimilarities to 
map distances using non-linear functions \citep{Sammon:1969}. More recently, other fields, like computer science have 
contributed methodological innovations, such as t-SNE \citep{van-der-Maaten+Hinton:2008}.

In the case of MDS, $D$ is a dissimilarity matrix with entries $\delta_{ij}$, and the 
fitted map distances—measured using the Euclidean norm  
\(d(x_i, x_j) = \|x_i - x_j\|_2\)—are matched to disparities \(\hat{\delta}_{ij}\), 
which are transformed versions of the input dissimilarities. The specific 
transformation \(\hat{\delta}_{ij} = f(\delta_{ij})\) depends on the MDS variant: 
in ratio MDS, \(f\) is a multiplicative transformation; in interval MDS, \(f\) is linear; 
and in non-metric MDS, \(f\) is restricted to monotonic functions.

The fit of an MDS model is quantified through the \emph{Stress} function

\begin{equation}\label{eq:stress}
  \mathrm{Stress}(X)\;=\;
  \sqrt{\frac{\displaystyle\sum_{i<j} \bigl(d(x_i, x_j) - \hat{\delta}_{ij}\bigr)^2}
               {\displaystyle\sum_{i<j} d(x_i, x_j)^2}}\,,
\end{equation}

% Optimization formulation
while the optimal configuration is obtained through

\begin{equation}\label{eq:xhat}
  \hat{X}\;=\;\arg\min_{X\in\mathbb{R}^{n\times d}}
  \mathrm{Stress}(X).
\end{equation}

The quality of $\hat{X}$ can be evaluated in multiple ways, including the Stress value, different agreement metrics 
like the Hit-Rate of K-nearest neighbor recovery \citep{Chen+Buja:2009}, or through visual comparisons, like  
Shepard Diagrams.  

\subsection{Dynamic mapping} \label{sec:dynamic-mapping}

Dynamic mapping extends static mapping by revealing how relationships among the $n$ objects change over time through
their movement paths in the lower-dimensional space. 

In dynamic mapping, the input data consist of a sequence of relationship matrices $(D_t)_{t=1, \dots, T}$
 rather than a single matrix. In general, these should be square matrices with a consistent set and ordering of 
objects across time. That is, each matrix $D_t \in \mathbb{R}^{n \times n}$ should represent relationships 
among the same $n$ objects.

Applying static mapping methods to each time point matrix $D_t$ independently introduces several challenges. 
The estimated point configurations $(\hat{X}_t)_{t=1, \dots, T}$, do 
often not align over time, preventing the analyst from tracking objects' movement. 
Further, results can be inconsistent due to local optima and are highly sensitive to 
variations in the data, including noise.

EvoMap, as proposed by \cite{Matthe+Ringel+Skiera:2023}, addresses these issues 
by incorporating static mapping methods into a dynamic optimization framework. 
EvoMap jointly fits a sequence of point configurations $(\hat{X}_t)_{t=1,\dots,T}$
to a sequence of dissimilarity matrices $(D_t)_{t=1,\dots,T}$, linking them over time 
through a regularization component. Thereby, it not only aligns subsequent configurations 
but also smooths the trajectories of objects over time. As a result, EvoMap 
produces coherent spatial representations of the evolving relationships 
 among the objects under analysis, which can be visualized in dynamic maps.

\begin{figure}[hbt!]
  \centering
  \includegraphics{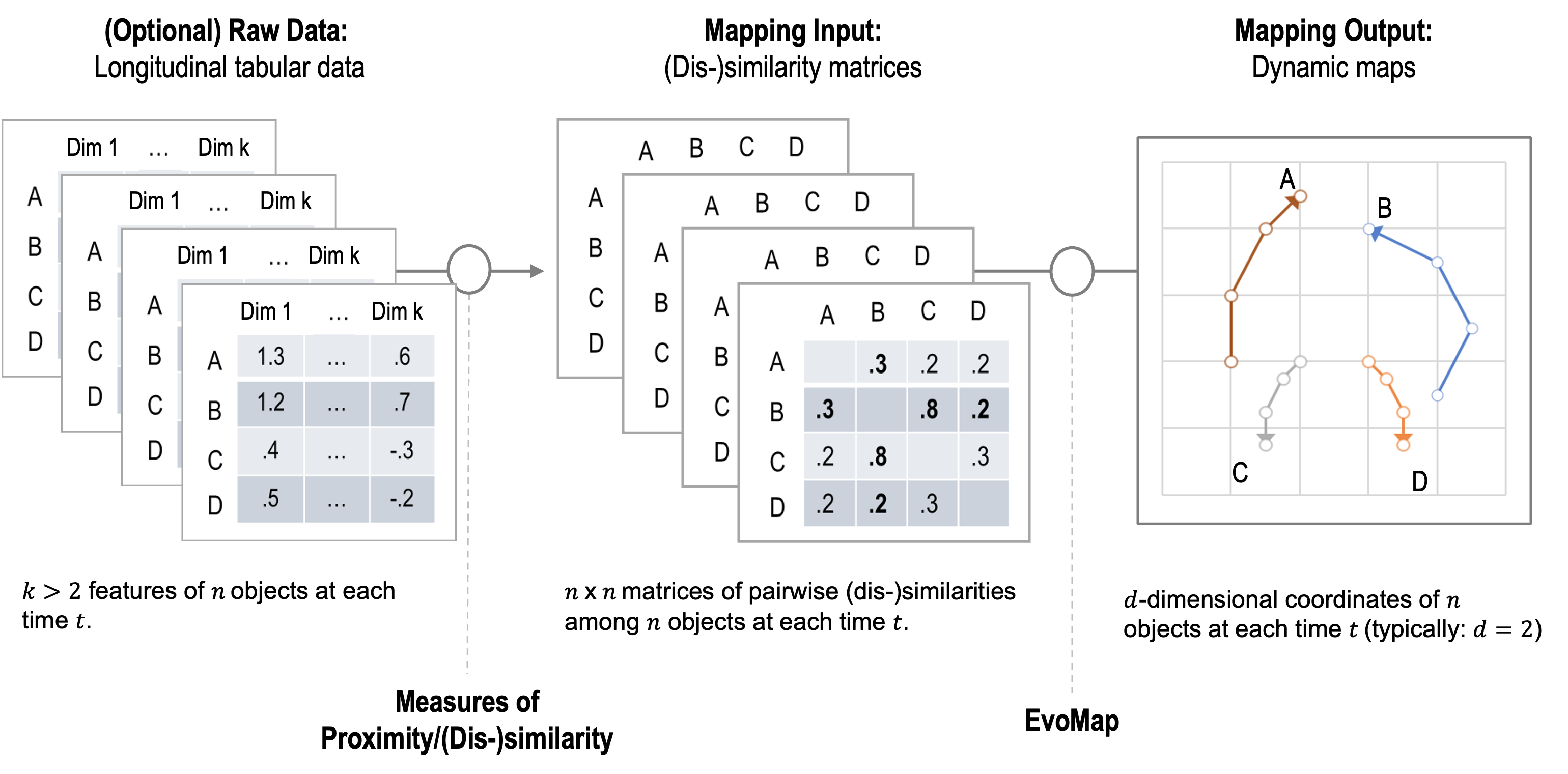}
  \caption{\label{fig:dynamic-mapping-overview} Illustration of dynamic mapping via EvoMap}
\end{figure}

Figure \ref{fig:dynamic-mapping-overview} illustrates the dynamic mapping process. 
Starting from longitudinal tabular data \((Z_t)_{t=1,\dots,T} \in \mathbb{R}^{n \times k}\), 
we derive a sequence of pairwise dissimilarity matrices \((D_t)_{t=1,\dots,T} \in \mathbb{R}^{n \times n}\) 
that quantify the relationships between objects at each time point. 
These matrices serve as input to EvoMap, which computes a sequence of configurations
\((\hat{X_t})_{t=1,\dots,T} \in \mathbb{R}^{n \times d}\) that preserve the higher-dimensional data's spatial structure 
as well as each object's temporal coherence in $d<k$ dimensions (typically, $d=2$). The first step may not be necessary if 
input data already consist of longitudinal relationship matrices, such as when analyzing temporal networks.

\FloatBarrier

Formally, EvoMap’s optimization problem is defined as:

\begin{equation} \label{eq:evomap-optim}
  (\hat{X}_t)_{t=1,\ldots,T} = \argmin_{X_1, \ldots, X_T \in \mathbb{R}^{n \times d}} C_{total}(X_1, \ldots, X_T),
\end{equation}

with the total cost function:

\begin{equation} \label{eq:cost-total}
  C_{total}(X_1, \ldots, X_T) = \sum_{t=1}^T{C_{static}(X_t)} + \alpha \cdot C_{temporal}(X_1, \ldots, X_T).
\end{equation}

Here, $C_{static}$ represents the static component of EvoMap’s cost function, which equals
the sum of the cost function $C_{static}$ of the selected static mapping method evaluated at each time 
$t \in \{1, \ldots, T\}$. When paired with MDS, for instance, $C_{static}$ would equal the Stress function in
Equation \ref{eq:stress}. $C_{temporal}$, weighted by the hyperparameter $\alpha \in \mathbb{R}$, represents the temporal component of the cost function, which
aligns the maps and smooths objects’ movements over time. 

Intuitively, when the hyperparameter $\alpha$ is set to zero, Equation \ref{eq:cost-total} 
corresponds to fully independent static mapping, where each configuration is fitted independently to the respective dissimilarity matrix $D_t$.
Setting $\alpha$ to a positive value introduces a penalty term, $C_{temporal}$, which penalizes objects for large changes in their
positions. As a result, the resulting configurations are aligned with each other, and the objects’ movement paths are smoothed.

Formally, the temporal cost component, $C_{temporal}$, is defined as:

\begin{equation} \label{eq:cost-temporal}
  C_{\text{temporal}}(X_1, \ldots, X_T) = 
  \sum_{i=1}^n f_w(i) \sum_{k=1}^p \sum_{t=k+1}^T  
   \left\| \nabla^k x_{i,t} \right\|^2
\end{equation}

where \\
\begin{description}
  \item $\nabla^k x_{i,t} \in \mathbb{R}^d$ is the $k$-th (backward) difference of object 
  $i$’s position at time $t$: \newline
  $\nabla^k x_{i,t} := \nabla^{k-1} x_{i,t}-\nabla^{k-1} x_{i,t-1}$, with 
  $\nabla^0 x_{i,t} := x_{i,t}$,

  \item $\left\| \nabla^k x_{i,t} \right\|^2$ denotes its squared Euclidean norm,

  \item $p \in \mathbb{N}^+$ is a second hyperparameter of positive integer values that 
  controls the degree of smoothing, and 

  \item $f_w : I \rightarrow \mathbb{R}^+$ assigns a positive weight for the temporal penalty to each object based on the degree of change
  in the input data.
\end{description}

Intuitively, the temporal cost function penalizes large movements of objects on the resultant configurations over time, 
indicated by large values of $\left\| \nabla^k x_{i,t} \right\|^2$. For $k=1$, this corresponds to the change in an object's position, 
measured as the squared Euclidean distance between the object's positions at time $t$ and $t-1$, summed over time. 
If the hyperparameter $p$ is larger than one, the regularization term also penalizes higher-order differences, i.e., 
changes in changes in positions. Ultimately, this penalizes larger, and in the case of $p>1$, also more volatile movements 
of objects over time. 

The object-specific weights $f_w(i)$ adjust the relative weight of this penalty for each object 
based on the observed changes in the input data. It is given by 

\begin{equation} \label{eq:weight-function}
f_w(i) := \exp(-b z_i)
\end{equation}

where $z_i := \sum_{t=2}^T \| D_{i,t} - D_{i,t-1} \|^2$ measures the total change in object $i$'s relationships through the 
change in the input matrices' corresponding row vectors $D_{i,t}$, and $b := \frac{1}{\max_{i \in I} z_i}$ is a normalizing constant.

For more details and intuition on EvoMap's cost function, we refer to \cite{Matthe+Ringel+Skiera:2023}. Note that 
the general form of this cost function assumes that there is a consistent global set of objects, and that all matrices
in the sequence retain their shape over time. Later, Section~\ref{sec:unbalanced_data} describes how EvoMap 
can generalize to unbalanced data, i.e., cases where the set of objects varies across time. 

The remainder of this paper details how to apply EvoMap using the \pkg{evomap} package.

\section[The evomap package]{The \pkg{evomap} package} \label{sec:package}

\subsection{Installation}

\pkg{evomap} can be installed directly from the Python Package Index (PyPI) by running:

\begin{CodeChunk}
  \begin{CodeInput}
  > pip install evomap
  \end{CodeInput}
\end{CodeChunk}

The installation requires \proglang{Python} version 3.9 or newer. 
At the time of writing the latest release of \pkg{evomap} is version 0.5.1. Updates are actively developed on GitHub, 
available at \href{https://github.com/mpmatthe/evomap}{https://github.com/mpmatthe/evomap} and subsequently released to PyPI.

\subsection{Key dependencies}

\pkg{evomap} relies on a limited number of other packages. It uses \pkg{numpy} and \pkg{scipy} for numerical 
computations, and \pkg{matplotlib} and \pkg{seaborn} for plotting. Additionally, \pkg{evomap} uses individual 
routines from \pkg{scikit-learn}, \pkg{scikit-optimize}, and \pkg{statsmodels}, and employs the just-in-time \proglang{C}-compiler \pkg{numba} 
to optimize runtime.

\subsection{Package overview}

\pkg{evomap} is structured into six core modules\footnote{For clarity, the unified term 'module' encompasses 
submodules, modules, and subpackages.}, designed to streamline the analysis workflow. 

Table \ref{tab:modules} presents an overview of these modules, with detailed descriptions and a complete API
reference available at \href{https://evomap.readthedocs.io/en/latest/}{\texttt{Read The Docs}}.

\vbox{%
\begin{enumerate}
  \item \code{datasets}: Offers example datasets, enabling users to familiarize themselves with the package's %
  capabilities. 
  \item \code{preprocessing}: Transforms input data into a format compatible with EvoMap. 
  \item \code{mapping}: Contains EvoMap implementations for various mapping methods.
  \item \code{printer}: Provides tools for visual exploration of both static and dynamic maps. 
  \item \code{transform}: Includes options for adjusting maps (e.g., rotations and reflections that can enhance %
  interpretability). 
  \item \code{metrics}: Features a set of evaluation metrics and goodness-of-fit statistics to evaluate mapping %
  results. 
\end{enumerate}}

\begin{table}[hbt!]
  \centering
  \small
  \renewcommand{\arraystretch}{1.1}
  \begin{tabular}{p{.15cm}p{2.15cm}p{3.75cm}p{3.25cm}p{3.5cm}}
  \hline
  \# & Module & Purpose & Example Function or Class & Description \\ 
  \hline
  (1) & \texttt{datasets} & Load sample datasets & 
  \code{load\_tnic\_snapshot()} & Load TNIC industry data \\

  (2) & \texttt{preprocessing} & Prepare input data & 
  \code{coocc2sim()} & Convert co-occurrence counts to similarity matrix \\
  & & & \code{edgelist2matrix()} & Convert edge list to adjacency matrix \\ 

  (3) & \texttt{mapping} & Fit EvoMap or baselines to data & 
  \code{MDS()}, \code{EvoMDS()} & Apply static or EvoMap-based MDS \\
  & & & \code{TSNE()}, \code{EvoTSNE()} & Apply static or EvoMap-based t-SNE \\

  (4) & \texttt{printer} & Visualize mapping results & 
  \code{draw\_map()} & Plot a single static map \\
  & & & \code{draw\_dynamic\_map()} & Plot overlay of multiple maps over time \\

  (5) & \texttt{transform} & Adjust mapping results & 
  \code{align\_maps()} & Align maps via Procrustes Analysis \\

  (6) & \texttt{metrics} & Evaluate mapping quality & 
  \code{misalign\_score()} & Measure temporal misalignment \\
  & & & \code{hitrate\_score()} & Measure nearest-neighbor recovery for static map \\
  & & & \code{persistence\_score()} & Measure temporal smoothness \\
  \hline
  \end{tabular}
  \caption{\label{tab:modules} Overview of key \texttt{evomap} modules and their purposes}
\end{table}

\FloatBarrier

\subsection{Package design and implementation}

\begin{figure}[hbt!]
  \centering
  \includegraphics{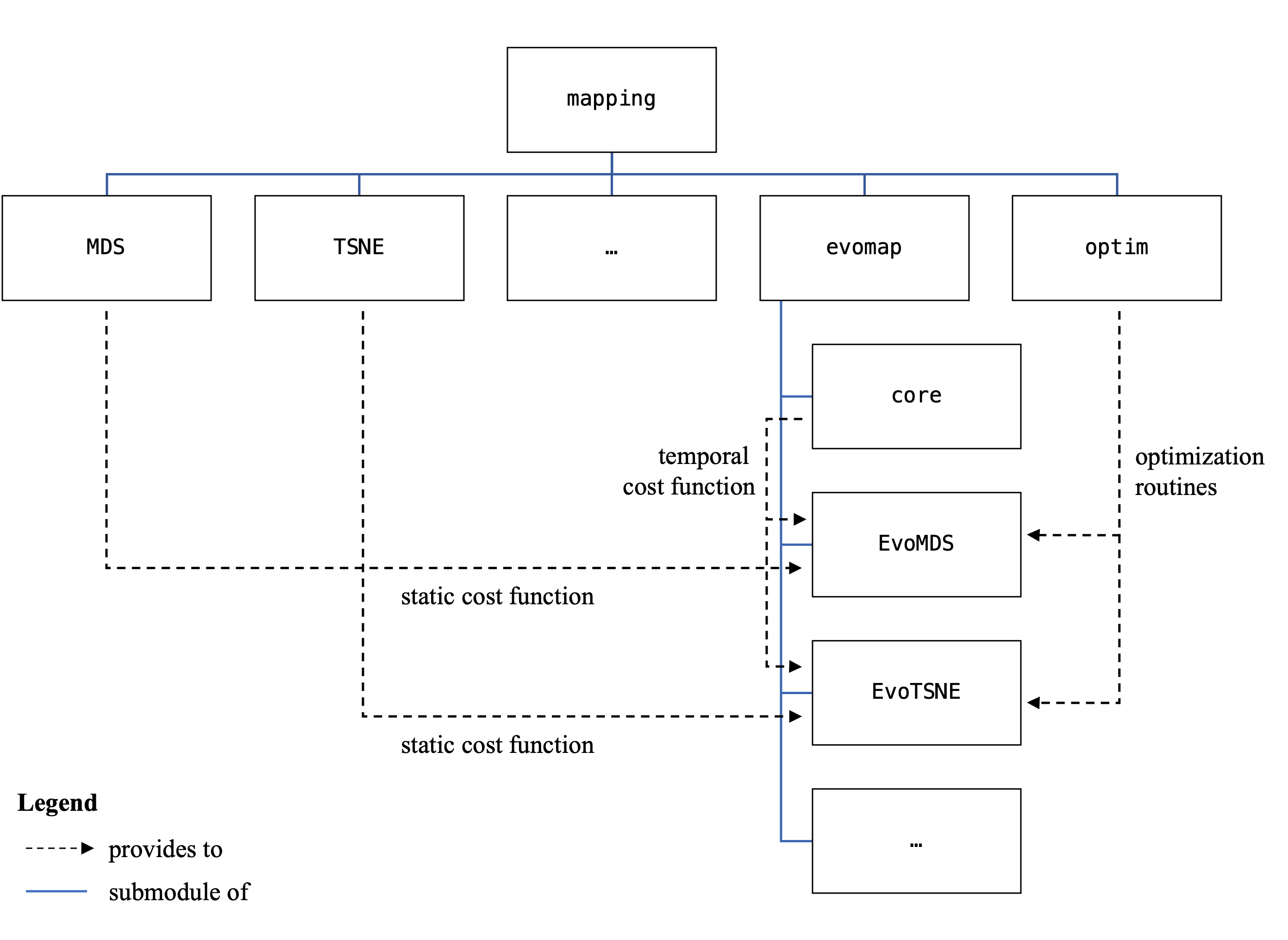}
  \caption{\label{fig:package-design} Overview of the mapping module}
\end{figure}

The \pkg{evomap} package is centered around its \code{mapping} module. This module is the heart
of the package and implements the EvoMap framework for various mapping methods along with 
their optimization routines. Figure~\ref{fig:package-design} illustrates its structure, which is designed with clear 
separation between shared and method-specific components.

At its center, the \code{core} module defines the abstract \code{EvoMap} class and implements the temporal cost 
function common to all implementations. The \code{optim} module provides optimization routines such as gradient descent 
with line search or momentum. Together, these modules centralize all functionalities that are used across implementations. 

In contrast, method-specific functionalities---most importantly, the static cost function---are provided by dedicated 
static classes and passed into their dynamic counterparts. As a result, extending the framework is easy, and all 
enhancements to the \code{core} framework affect all mapping implementations uniformly.

As described in Section~\ref{sec:dynamic-mapping}, EvoMap’s cost function contains a
static and a temporal component--the former varies by mapping method, while the latter
is consistent across all EvoMap implementations. Accordingly, the mapping module follows
a dual structure as well. There, individual static implementations of each mapping method provide
their respective static cost components, while a dynamic implementation integrates the method into the
EvoMap framework. For MDS, for instance, the mapping module includes both \code{MDS}, which provides the 
static cost function for Multidimensional Scaling, while \code{EvoMDS} integrates this method into the 
dynamic EvoMap framework.

To illustrate, consider the implementation of \code{EvoMDS}. This class inherits from \code{EvoMap} to draw on  
the shared temporal cost function. It receives its static cost function from the separate \code{MDS} class, which 
implements \code{\_normalized\_stress\_function}. The \code{fit\_transform()} method in 
\code{EvoMDS} then combines both cost components and passes them to the optimizer. Its structure closely mirrors 
other implementations like \code{EvoSAMMON} or \code{EvoTSNE}, differing mainly in the static component and the choice 
of the optimizer.

Thanks to this structure, extending \pkg{evomap} with a new mapping method is straightforward: 
(i) define a new static class implementing the method’s cost function (e.g., as in \code{MDS}), and 
(ii) create a subclass of \code{EvoMap} that integrates this function into the dynamic framework 
by overriding \code{fit\_transform()} (e.g., as in \code{EvoMDS}). 

\FloatBarrier

Currently, \pkg{evomap} includes dynamic implementations for three methods---MDS, Sammon Mapping, and t-SNE---which 
can serve as blueprints for future extensions. The next section illustrates how to apply these methods in practice.

Note that the results reported in this paper were generated using Python version 3.13.2 and \pkg{evomap} 
version 0.5.1 on macOS 15.1.1. While the \pkg{evomap} package is designed to be platform-independent, 
minor differences in results may occur across operating systems (e.g., due to variations in floating-point 
precision). However, results on other systems are expected to be very similar, and any differences should not 
affect the qualitative conclusions drawn from the analyses.

%% -- Detailed Usage Example -------------------------------------------

\section{Detailed usage example} \label{sec:usage-example}

This section demonstrates how to apply the \pkg{evomap} package to a subset of the 
Text-based Network Industries Classification (TNIC) dataset by \cite{Hoberg+Phillips:2016}. 
It provides a step-by-step guide and showcases the tools within \pkg{evomap} that facilitate 
each phase of the analysis.

The TNIC dataset captures firm–firm relationships based on the linguistic similarity of 
product descriptions in annual reports. The data consist of a time-indexed edgelist, 
where each row represents a pair of firms at a given time, including their names, a 
similarity score, and a standardized identifier that can be used to link firms to additional 
information (e.g., their Standard Industrial Classification (SIC) codes or market size). 
Intuitively, the TNIC similarity scores indicate product-market similarity—that is, the degree to 
which two firms compete by offering related products or services. As such, TNIC similarity 
scores can be interpreted as measures of competitive proximity, which—through mapping 
techniques—can reveal firms' underlying spatial positions in the market. For further 
information on the TNIC dataset, we refer to \cite{Hoberg+Phillips:2016}.

We begin by importing the necessary libraries and fixing the random seed:
\begin{Code}
>>> import numpy as np
>>> import pandas as pd
>>> from matplotlib import pyplot as plt
>>> np.random.seed(123)
\end{Code}

A subset of the TNIC data is available as an example through the \code{datasets} module:

\begin{Code}
>>> from evomap.datasets import load_tnic_sample_tech
>>> data = load_tnic_sample_tech()
\end{Code}

The subset used for this illustration focuses on nine technology 
companies—Apple, AT\&T, eBay, Intuit, Micron Technology, Microsoft, Oracle, US Cellular, and Western 
Digital—spanning twenty years from 1998 to 2017. Each data point represents a pair of firms at a 
given time and includes their names, a similarity score, Standard Industrial Classification (SIC) 
codes, and a size variable derived from market values.

To better understand the structure of the input data, Table~\ref{tab:data-overview} lists the first 
and last five rows of the dataset. Each row records the year, firm names (\texttt{name1}, \texttt{name2}), 
their pairwise similarity score, SIC codes (\texttt{sic1}, \texttt{sic2}), and size variables 
(\texttt{size1}, \texttt{size2}).

 \begin{Code}
>>> table_overview = pd.concat([data.head(), data.tail()], axis=0)
>>> table_overview = table_overview[
...     ['year', 'name1', 'name2', 'score', 'sic1', 'sic2', 'size1', 'size2']
... ].round(2)
\end{Code}

\begin{table}[t!]
  \centering
  \begin{tabular}{cp{3cm}p{3cm}cccccc}
  \hline
  year & name1 & name2 & score & sic1 & sic2 & size1 & size2 \\ 
  \hline
  1998 & APPLE INC & WESTERN DIGITAL CORP & 0.07 & 36 & 35 & 71.79 & 32.29 \\
  1998 & APPLE INC & MICROSOFT CORP & 0.06 & 36 & 73 & 71.79 & 517.38 \\
  1998 & APPLE INC & ORACLE CORP & 0.04 & 36 & 73 & 71.79 & 188.44 \\
  1998 & AT\&T INC & US CELLULAR CORP & 0.08 & 48 & 48 & 324.14 & 57.62 \\
  1998 & EBAY INC & MICROSOFT CORP & 0.03 & 73 & 73 & 98.54 & 517.38 \\
  2017 & ORACLE CORP & MICROSOFT CORP & 0.13 & 73 & 73 & 432.13 & 728.91 \\
  2017 & ORACLE CORP & INTUIT INC & 0.02 & 73 & 73 & 432.13 & 187.3 \\
  2017 & US CELLULAR CORP & AT\&T INC & 0.02 & 48 & 48 & 56.56 & 488.57 \\
  2017 & WESTERN DIGITAL CORP & APPLE INC & 0.03 & 35 & 36 & 161.4 & 888.85 \\
  2017 & WESTERN DIGITAL CORP & MICRON TECHNOLOGY INC & 0.08 & 35 & 36 & 161.4 & 188.55 \\
  \hline
  \end{tabular}
  \caption{Overview of the TNIC sample data} \label{tab:data-overview}
  \begin{flushleft}
    {\footnotesize \textit{Note:} The data represent an edgelist of pairwise observations. \texttt{name1} and \texttt{name2} refer to the first and second firm in each row, \texttt{sic1} and \texttt{sic2} to their respective SIC codes, and \texttt{size1} and \texttt{size2} to their firm sizes.}
  \end{flushleft}
  \end{table}

\FloatBarrier

\subsection{Preprocessing}

The \pkg{evomap} package is designed to analyze longitudinal relationship data, which must be provided 
as a sequence of $n \times n$ relationship matrices, where $n$ is the number of objects. Many mapping 
methods, including MDS, require dissimilarities as input. When the available data do not directly meet 
this requirement, \pkg{evomap} provides preprocessing functions to convert different data formats into 
a suitable form.

In the TNIC example, the data are initially given as a time-indexed edgelist that captures similarities, 
not dissimilarities. The first preprocessing step, therefore, involves converting the edgelist into a 
sequence of square matrices. This is accomplished using the \code{edgelist2matrices()} function, 
where the user specifies the columns corresponding to object identifiers, similarity scores, and time.

\begin{Code}
>>> from evomap.preprocessing import edgelist2matrices
>>> S_t, labels_t = edgelist2matrices(
...     data,
...     score_var='score',
...     id_var_i='name1',
...     id_var_j='name2',
...     time_var='year')
\end{Code}

The output, \code{S\_t}, now contains the data in the format required by \pkg{evomap}: a sequence of 
square matrices, while \code{labels\_t} indicates the object names corresponding to the rows and 
columns of each matrix.

The next step involves converting the similarity measures in \code{S\_t} into dissimilarities. 
In the TNIC dataset, the input values are pairwise similarities $s_{ij} \in [0,1]$, where higher 
values indicate greater similarity between firms. To derive dissimilarities, we apply a simple 
transformation, $\delta_{ij} = 1 - s_{ij}$, which maps high similarity to low dissimilarity and 
vice versa. This transformation—along with additional options for unbounded similarities—is 
implemented in the \code{sim2diss()} function.

Applying this transformation to each matrix in the sequence yields a list of 20 dissimilarity 
matrices $(D_t)_{t = 1, \dots, 20}$, each of shape $(9, 9)$, ready for dynamic mapping.

\begin{Code}
>>> from evomap.preprocessing import sim2diss
>>> D_t = [sim2diss(S, transformation='mirror') for S in S_t]
\end{Code}

\subsection{Fitting EvoMap to data}
The general process for fitting EvoMap to data is straightforward and involves the following three steps:

\begin{enumerate}
  \item \textbf{Import the \code{EvoMap} class} corresponding to the desired mapping method—e.g., 
  \code{EvoMDS} for non-metric or metric MDS, \code{EvoSAMMON} for Sammon mapping, or 
  \code{EvoTSNE} for t-SNE.
  
  \item \textbf{Initialize EvoMap} with the appropriate parameters, including the hyperparameters 
  \code{alpha} and \code{p}, as well as method-specific arguments (e.g., \code{mds_type} for MDS or \\ 
  \code{perplexity} for t-SNE).
  
  \item \textbf{Fit EvoMap} to the input data using the \code{fit\_transform()} method.
\end{enumerate}

In the TNIC example, the input consists of a sequence of 20 dissimilarity matrices for 9 firms, 
each stored as a \pkg{numpy} array of shape \code{(9, 9)}.

\begin{CodeChunk}
\begin{CodeInput}
>>> print(len(D_t))       # Output: 20
>>> print(D_t[0].shape)   # Output: (9,9)
\end{CodeInput}
\end{CodeChunk}

To apply non-metric MDS to this sequence of matrices, we first import and instantiate 
\code{EvoMDS()}. We specify the \code{mds\_type} as \code{'ordinal'} to select the non-metric 
variant, set the hyperparameters $\alpha$ and $p$ (whose roles are discussed later in more detail), and define the 
number of random initializations using the \code{n\_inits} argument.

\begin{Code}
>>> from evomap.mapping import EvoMDS
>>> evomds = EvoMDS(alpha=0.4, p=2, mds_type='ordinal', n_inits=10)
\end{Code}

Then, we fit EvoMap to the data: 

\begin{Code}              
>>> X_t = evomds.fit_transform(D_t)
\end{Code}

The result, \code{X_t}, is a sequence of \code{numpy} arrays,  
positioning the firms on two dimensions over 20 years. 

\begin{CodeChunk}
\begin{CodeInput}
>>> print(len(X_t))       # Output: 20
>>> print(X_t[0].shape)   # Output: (9,2)   
\end{CodeInput}
\end{CodeChunk}

\begin{figure}
  \centering
  \resizebox{1\textwidth}{!}{\includegraphics{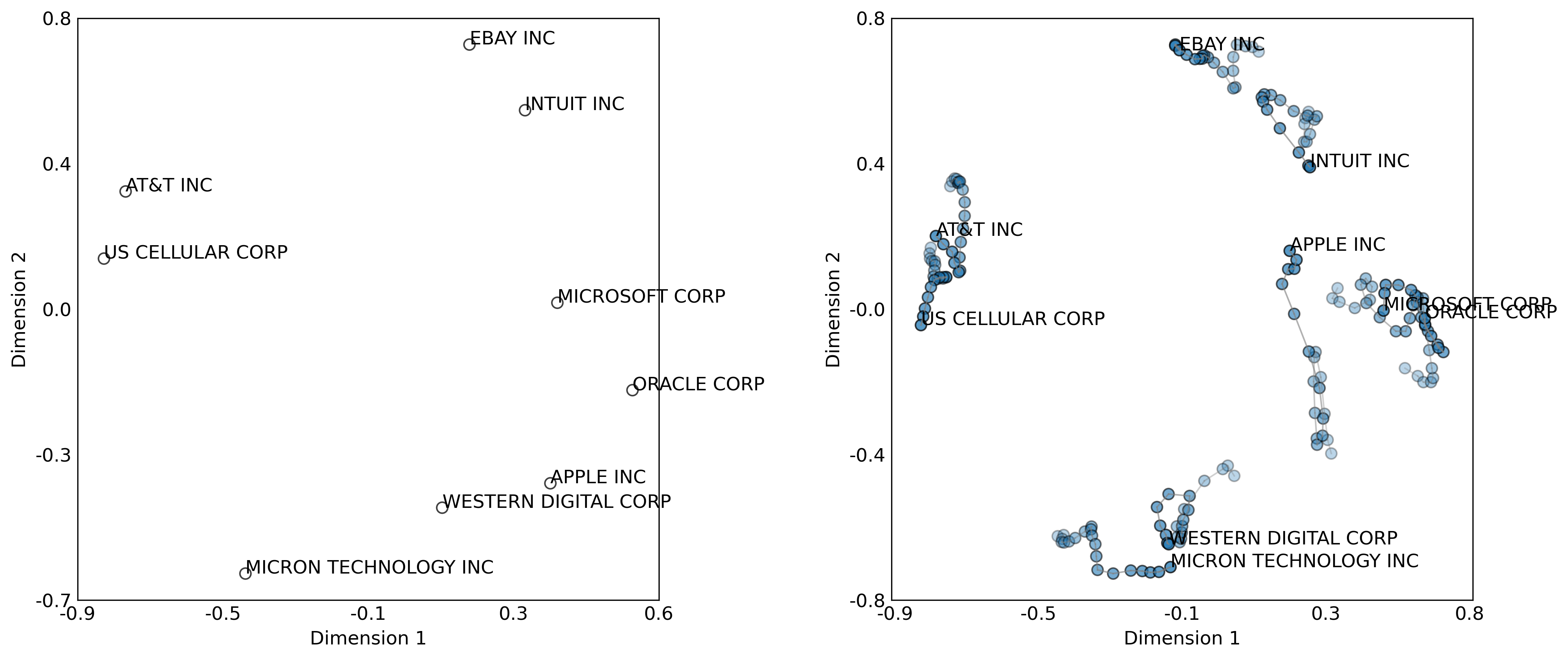}}
  \caption{\label{fig:dynamic-map} Dynamic map of nine technology firms, 1998-2017}
\end{figure}

To explore the result, we can visualize the results with \code{draw_dynamic_map()},
providing the point coordinates, labels, and optionally displaying movement arrows and axes. 
Figure \ref{fig:dynamic-map} showcases the nine firms' initial positions in 1998 (left) and the subsequent 
movement paths over time until 2017 (right).  

\begin{CodeChunk}
\begin{CodeInput}
>>> from evomap.printer import draw_map, draw_dynamic_map
>>> fig, ax = plt.subplots(1, 2, figsize=(14, 6))
>>> draw_map(
...     X_t[0], 
...     label=labels_t[0],
...     show_axes=True,
...     ax=ax[0])
>>> draw_dynamic_map(
...     X_t,
...     label=labels_t[0],
...     show_arrows=True,
...     show_axes=True,
...     transparency_start=0.3,
...     transparency_end=0.7, 
...     ax=ax[1])
\end{CodeInput}
\end{CodeChunk}

\FloatBarrier

The static snapshot (left) reveals that, within the nine depicted firms, there are four groups 
of similar firms based on their main products sold in 1998: i) hardware-focused firms, such 
as Apple, Western Digital, and Micron Technology; ii) platform-focused firms, like Microsoft and Oracle, 
iii) software-focused firms, such as eBay and Intuit, and 
iv) telecommunication service providers, like AT\&T and US Cellular. Among these pairs, 
the telecommunication firms appear more distinct compared to the remaining groups
whose products commonly relate to computers. Within the three computer-related groups,
Microsoft and Oracle, known for platforms like the Windows Operating System or Oracle’s
Database System, appear between software and hardware providers. Among the hardware
providers, Micron Technologies, primarily known as a semiconductor firm, is differentiated
from Western Digital, known for storage solutions, and Apple, known for personal computers
like the iMac. 

The dynamic map on the right highlights key shifts in the firms’ positions:
Apple and Western Digital diverge as Apple ventures into the software space, drawing it
closer to firms like Intuit. Meanwhile, Microsoft and Oracle--both heavily invested in cloud services--converge, 
and the positions of telecommunication firms remain relatively static.

\FloatBarrier

\subsection{Detailed setup of EvoMap}

When using EvoMap, the analyst must make several parameter choices. The key parameters to consider when initializing EvoMap include:
\begin{itemize}
  \item \textbf{Starting positions}, which can be provided via the \code{init} argument to initialize EvoMap with %
  fixed, rather than random, starting positions. 
  \item \textbf{EvoMap's hyperparameters}, $\alpha$ and $p$, which control the degree of alignment and smoothness. %
  \item \textbf{Optimization arguments}, which impact convergence and computational time. For example, when using %
  random, rather than fixed, starting positions, \code{n_inits} sets the number of different random initializations %
   for the optimization algorithm.  
  \item \textbf{Method-specific arguments}, which vary by the mapping method used. For instance, MDS requires % 
  specifying the \code{mds_type} (e.g., ratio, interval, or ordinal), while t-SNE requires setting its \code{perplexity}. 
\end{itemize}

The following sections outline some of these choices in more detail. 

\subsubsection{Starting positions}

Optimization-based mapping methods can be sensitive to the choice of starting positions. 
To control this aspect, EvoMap provides two main options: specifying a number of random 
initializations via \code{n\_inits}, or providing fixed starting positions via the \code{init} 
argument.

The first option involves running the model multiple times with different random 
starting positions. This can help increase robustness by reducing the risk of getting stuck 
in local minima. The number of runs is controlled via the \code{n\_inits} argument, which 
accepts a positive integer. When \code{n\_inits} is greater than one, EvoMap runs the 
optimization multiple times and retains the solution with the lowest cost. To ensure 
reproducibility in this case, users can fix the random seed beforehand 
(e.g., via \code{np.random.seed()}).

Alternatively, users can supply explicit starting positions through the \code{init} argument. 
This option provides full control over initialization and can often be useful to improve convergence by 
starting from a sensible configuration—such as the output of a 
static method like Classical Scaling \citep{DeLeeuw+Mair:2009}, which is available as \code{evomap.mapping.CMDS}. 
The \code{init} argument expects a list of coordinates matching the desired output shape—in this case, \code{(9, 2)}. 

\textbf{Note:} If \code{init} is set, \code{n\_inits} is ignored and only a single run is performed.

In the example above, we employ the first strategy and initialize the starting positions with ten different 
random draws by setting \code{n_inits=10}. The final solution stored in \code{X_t} retains the one corresponding to 
the lowest cost function value. 

\subsubsection{EvoMap's hyperparameters}

EvoMap requires the user to set two hyperparameters: $\alpha$ and $p$. 
$\alpha$ influences how strongly subsequent point coordinates align with each other, 
whereas $p$ controls the smoothness of each object's movement path across time. Carefully selecting
these hyperparameters is crucial, as they strongly affect EvoMap's output. 

Users of the \pkg{evomap} package can identify suitable values for $\alpha$ and $p$ in two primary ways: 

\begin{enumerate}
  \item \textbf{Manual inspection}, by exploring EvoMap's output under various hyperparameter combinations. 
  \item \textbf{Automated evaluation}, by using a grid search or Bayesian optimization to systematically test and evaluate a range of %
  hyperparameter combinations.
\end{enumerate}

To illustrate the influence of EvoMap's hyperparameters, we examine the effects of different $\alpha$ and $p$ 
values on EvoMap's output using the TNIC dataset as an example. 

First, we generate a set of different hyperparameter values:

\begin{CodeChunk}
\begin{CodeInput}
>>> configs = [
...     {"alpha": 0.01, "p": 1}, 
...     {"alpha": 0.4,  "p": 1}, 
...     {"alpha": 1.5,  "p": 1}, 
...     {"alpha": 0.4,  "p": 1}, 
...     {"alpha": 0.4,  "p": 2}, 
...     {"alpha": 0.4,  "p": 3}]
\end{CodeInput}
\end{CodeChunk}

Then, we fit EvoMap using these values. To make the resultant configurations easily comparable, we provide fixed starting positions obtained by Classical Scaling.

\begin{CodeChunk}
\begin{CodeInput}
>>> from evomap.mapping import CMDS
>>> cmds_t = [CMDS().fit_transform(D_t[0]) for D in D_t]
>>> solutions = [EvoMDS(**cfg, mds_type='ordinal', init=cmds_t)
...               .fit_transform(D_t) for cfg in configs]
\end{CodeInput}
\end{CodeChunk}

Finally, we plot the resultant configurations to compare the results visually.

\begin{CodeChunk}
\begin{CodeInput}
>>> titles = ['A: Low alpha / p=1', 'B: Medium alpha / p=1', 
...           'C: High alpha / p=1', 'D: Medium alpha / p=1',
...           'E: Medium alpha / p=2', 'F: Medium alpha / p=3']
>>> fig, ax = plt.subplots(3, 2, figsize=(14, 21))
>>> for i, (X, title) in enumerate(zip(solutions, titles)):
...     row, col = i % 3, i // 3
...     draw_dynamic_map(X,
...                      label=labels_t[0],
...                      show_axes=True,
...                      show_arrows=True,
...                      show_last_positions_only=True,
...                      ax=ax[row, col])
...     ax[row, col].set_title(f'Solution {title}', fontsize=12)
...     ax[row, col].set_xlabel('Dimension 1')
...     ax[row, col].set_ylabel('Dimension 2')
\end{CodeInput}
\end{CodeChunk}

\begin{figure}[hbt!]
  \centering
  \resizebox{\textwidth}{!}{\includegraphics{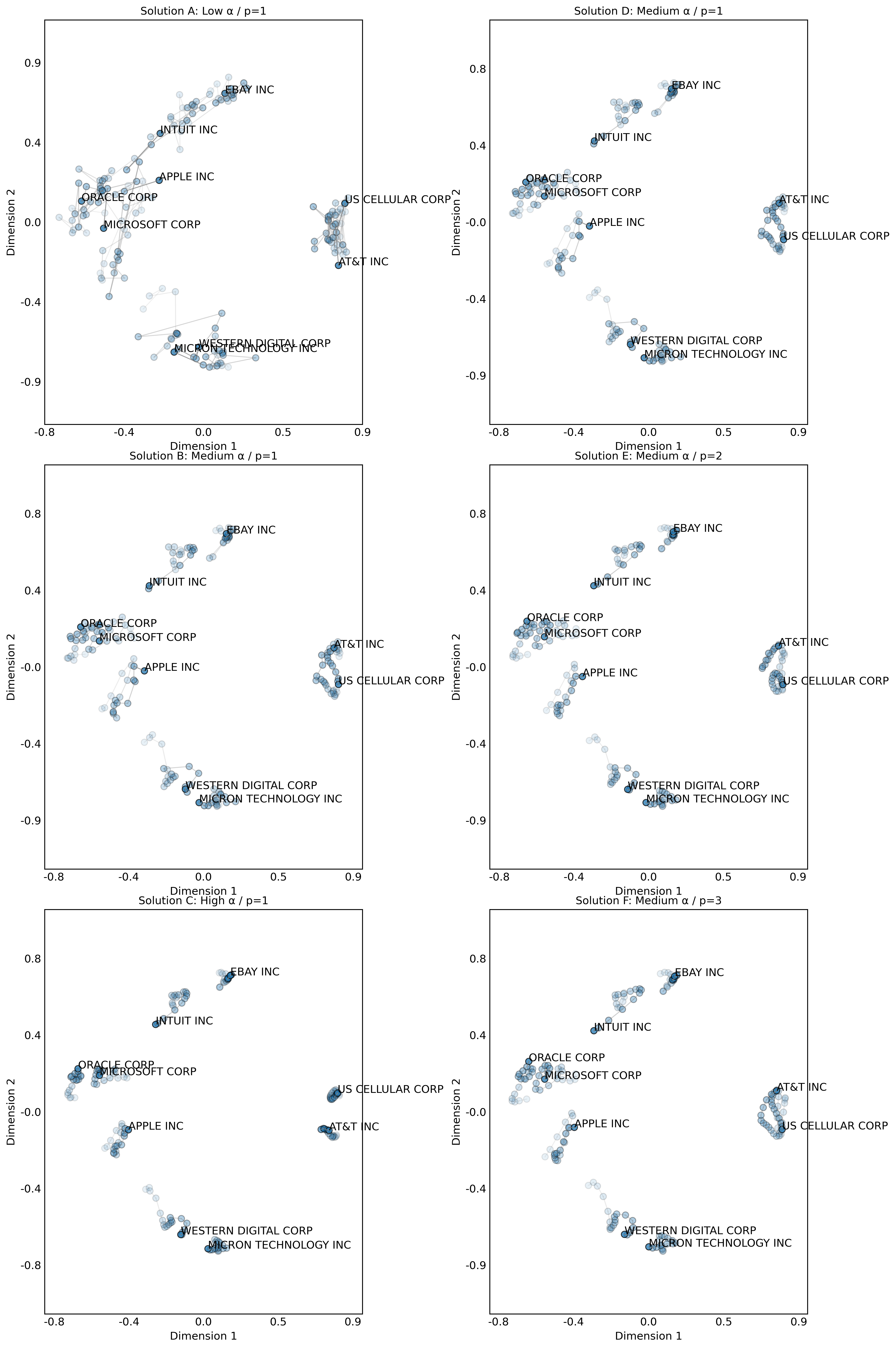}}
  \caption{\label{fig:hyperparameters} Dynamic map under different hyperparameter choices}
\end{figure}

Figure \ref{fig:hyperparameters} displays the resulting configurations for the selected hyperparameter values and 
illustrates how each hyperparameter influences the mapping results. The first column varies $\alpha$ (with $p$ 
held constant at 1), while the second column varies $p$ (with $\alpha$ fixed at a medium value).

As shown in Figure \ref{fig:hyperparameters}, lower values of $\alpha$ tend to yield maps with weaker alignment 
and more erratic movement paths. As $\alpha$ increases (from top to bottom in the first column), the maps become 
more aligned and the movement paths shorter. Extremely high values of $\alpha$ can result in nearly static maps 
with only minimal temporal change. In contrast, the effect of $p$ is less about path length and more about path 
smoothness: higher $p$ values reduce zigzag movements, producing smoother trajectories and more consistent 
distances between successive time points (see second column in Figure \ref{fig:hyperparameters}, top to bottom).

Section \ref{sec:hyperparameters} will discuss methodological approaches, grid search and Bayesian optimization, 
for exploring appropriate values for $\alpha$ and $p$.

\FloatBarrier

\subsubsection{Convergence diagnostics}

Fitting EvoMap operates quietly by default, providing no details about the convergence
of its gradient-based optimization. However, such information on convergence can be informative, 
especially when faced with suboptimal solution quality or unexpected results. 
To obtain such diagnostics, users can modify the \code{verbose} parameter to increase output verbosity: 
\begin{itemize}
  \item \code{verbose=0} (the default) means no output is provided.
  \item \code{verbose=1} reveals basic information about the optimization routine, the final cost function value, and % 
  convergence status.
  \item \code{verbose=2} offers detailed insights into the optimization process, including cost function values at %
  specific intervals.  
\end{itemize}

Setting \code{verbose=1}, for instance, indicates the optimization routine (gradient descent with backtracking)
and its successful convergence at iteration 189:

\begin{CodeChunk}
\begin{CodeInput}
>>> EvoMDS(
...     alpha=.4,
...     p=2,
...     mds_type='ordinal',
...     init=cmds_t,
...     verbose=1).fit(D_t)
\end{CodeInput}
\begin{CodeOutput}
[EvoMDS] Running Gradient Descent with Backtracking via Halving
[EvoMDS] Iteration 189: gradient norm vanished. Final cost: 3.94
\end{CodeOutput}
\end{CodeChunk}

Here, we use \code{fit()}, which runs EvoMap without returning the resultant point coordinates. 
Setting \code{verbose=2} and \code{n_iter_check=50} evaluates the cost function every 50 iterations and reveals 
how the cost function values decrease monotonically, until the gradient norm vanishes around iteration 189.

\begin{CodeChunk}
\begin{CodeInput}
>>> EvoMDS(
...     alpha=.4,
...     p=2,
...     mds_type='ordinal',
...     init=cmds_t,
...     n_iter_check=50,
...     verbose=2).fit(D_t)
\end{CodeInput}
\begin{CodeOutput}
[EvoMDS] Running Gradient Descent with Backtracking via Halving
[EvoMDS] Iteration 50 -- Cost: 3.99 -- Gradient Norm: 0.0539
[EvoMDS] Iteration 100 -- Cost: 3.95 -- Gradient Norm: 0.0080
[EvoMDS] Iteration 150 -- Cost: 3.94 -- Gradient Norm: 0.0048
[EvoMDS] Iteration 189: gradient norm vanished. Final cost: 3.94
\end{CodeOutput}
\end{CodeChunk}

Should these diagnostics indicate issues with convergence, several strategies can help improve optimization performance.
Specifically, the user can:

\vbox{
\begin{itemize}
  \item Utilize \code{evomap.preprocessing} to normalize input data, as some mapping methods struggle with input data measured on unusual or very large scales.
  \item Increase \code{n_iter} to extend the number of allowed iterations.
  \item Adjust \code{step_size} (by default: 1) to modify convergence speed; decrease if optimization overshoots, or increase it for faster convergence. 
  \item Experiment with multiple initializations through \code{n_inits} to mitigate the impact of random starting positions. 
\end{itemize}
}

Once the user is confident about proper convergence of the optimization, one can start exploring
the results in more depth and evaluate the solution quality more rigorously.

\subsection{Exploration}

The \pkg{evomap} package provides a range of options to explore mapping results visually through the \code{printer}
module. One can categorize these functions into two groups:

\begin{enumerate}
  \item \textbf{Static maps}, which visualize individual configurations at a time.
  \item \textbf{Dynamic maps}, which visualize their evolution over time.
\end{enumerate}

\subsubsection{Static maps}

\begin{figure}[hbt!]
\centering
\resizebox{\textwidth}{!}{\includegraphics{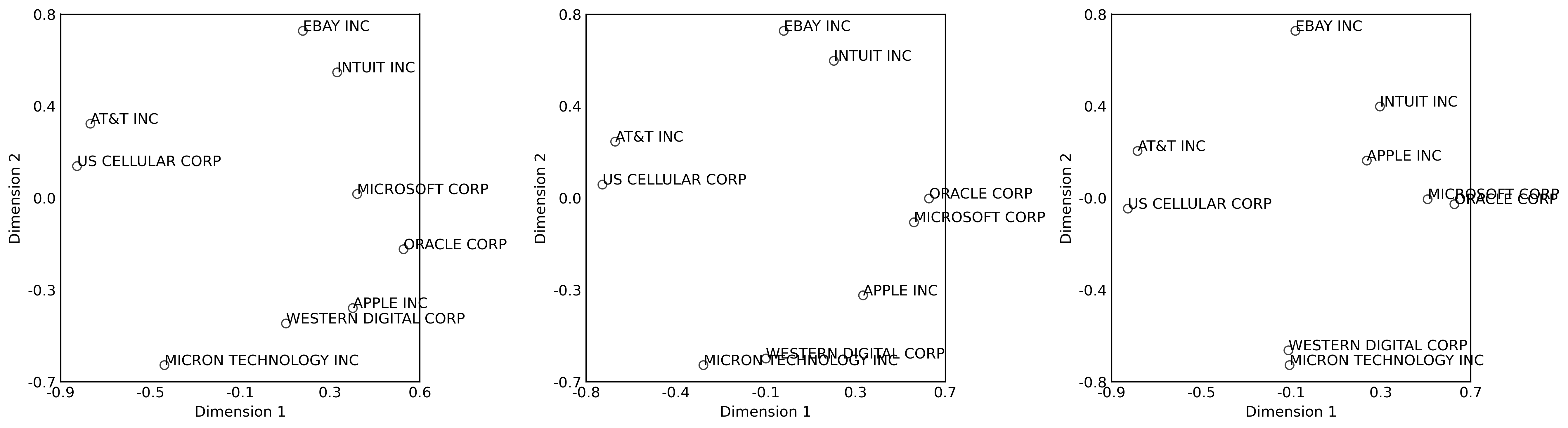}}
\caption{\label{fig:static-snapshots} Three static EvoMap snapshots}
\end{figure}

Static snapshots help evaluate the quality of mapping solutions and set reference points
for dynamic exploration. The primary function for static visualization is \code{draw_map()},
which, in its simplest form, requires just the point coordinates $\hat{X}$ to generate a scatter plot. 
For unidimensional data, it plots a line scale. Figure \ref{fig:static-snapshots} displays three examples of static 
snapshots.

\begin{CodeChunk}
\begin{CodeInput}
>>> fig, ax = plt.subplots(1, 3, figsize=(18, 5))
>>> labels = labels_t[0]
>>> draw_map(X_t[0], label=labels, ax=ax[0])
>>> draw_map(X_t[10], label=labels, ax=ax[1])
>>> draw_map(X_t[19], label=labels, ax=ax[2])
\end{CodeInput}
\end{CodeChunk}

Further customization is possible by linking additional object-specific information to visual aesthetics. In the TNIC 
example, the data include additional information about each firm's name, its SIC code, and its size. To bind these 
attributes to visual aesthetics of the maps, we first store them in individual lists.

\begin{CodeChunk}
\begin{CodeInput}
>>> sic_codes = [
...     data.query('name1 == @firm').sic1.unique()[0]
...     for firm in labels]
>>> sizes = [
...     data.query('name1 == @firm').size1.unique()[0].round(2)
...     for firm in labels]
>>> periods = data.year.unique()
\end{CodeInput}
\end{CodeChunk}
 
\begin{figure}[hbt!]
  \centering
  \resizebox{1\textwidth}{!}{\includegraphics{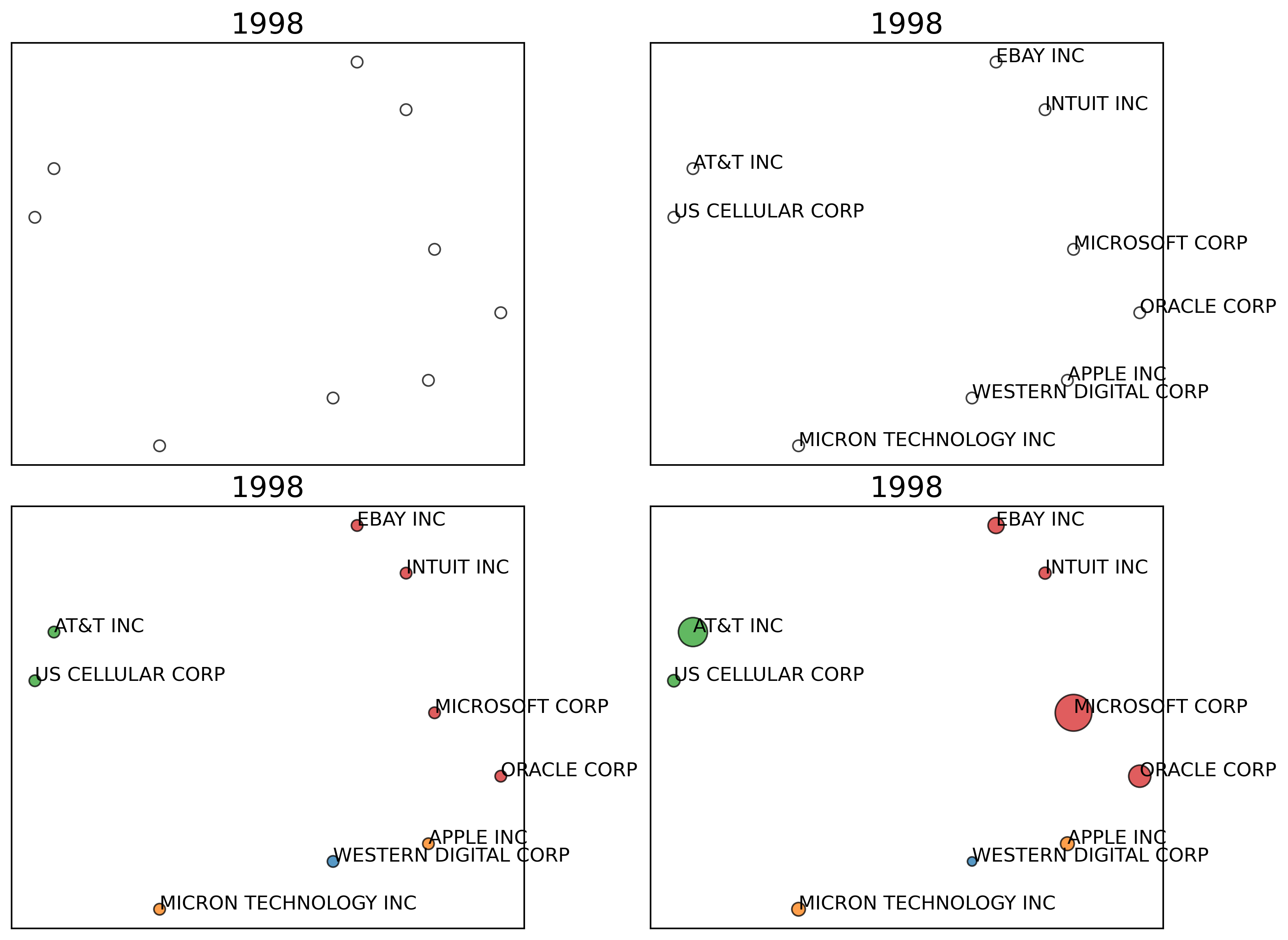}}
  \caption{\label{fig:draw-map-illustrations} Static map with different aesthetics}
\end{figure}

These lists can then be provided to the \code{label}, \code{color}, and \code{size} arguments. The results 
are shown in Figure \ref{fig:draw-map-illustrations}.

\vbox{
\begin{CodeChunk}
\begin{CodeInput}
>>> fig, ax = plt.subplots(2, 2, figsize=(11, 8), layout='constrained')
>>> draw_map(X_t[0],
...          show_axes=False,
...          title=periods[0],
...          ax=ax[0, 0])
>>> draw_map(X_t[0],
...          label=labels,
...          show_axes=False,
...          title=periods[0],
...          ax=ax[0, 1])
>>> draw_map(X_t[0],
...          label=labels,
...          color=sic_codes,
...          show_axes=False,
...          show_legend=False,
...          title=periods[0],
...          ax=ax[1, 0])
>>> draw_map(X_t[0],
...          label=labels,
...          color=sic_codes,
...          size=sizes,
...          show_axes=False,
...          title=periods[0],
...          ax=ax[1, 1])
\end{CodeInput}
\end{CodeChunk}
}

\FloatBarrier

\subsubsection{Dynamic maps}

Beyond individual snapshots, the package allows users to create dynamic maps, 
which overlay multiple snapshots to illustrate changes over time. 
This is accomplished with \code{draw_dynamic_map()} or \code{draw_trajectories()}.

Dynamic maps incorporate the same aesthetic customizations as static maps but extend them to reflect changes 
across time points. Thus, variables like color or size should be provided as lists of arrays, each containing
  all objects' attribute values at a specific time point: 

\begin{CodeChunk}
\begin{CodeInput}
>>> sic_codes_t = []
>>> sizes_t = []
>>> for t in range(len(X_t)):
...     data_t = data.query('year == @periods[@t]')
...     sic_codes_t.append(np.array([
...         data_t.query('name1 == @firm').sic1.unique()[0]
...         for firm in labels]))
...     sizes_t.append(np.array([
...         data_t.query('name1 == @firm').size1.unique()[0]
...         for firm in labels]))
\end{CodeInput}
\end{CodeChunk}

These lists can then be used as arguments for \code{draw_dynamic_map()}:

\begin{CodeChunk}
\begin{CodeInput}
>>> fig, ax = plt.subplots(1, 2, figsize=(16, 7))
>>> draw_dynamic_map(X_t,
...                  label=labels,
...                  color_t=sic_codes_t,
...                  size_t=sizes_t,
...                  show_arrows=True,
...                  show_axes=True,
...                  ax=ax[0])
\end{CodeInput}
\end{CodeChunk}

\begin{figure}[hbt!]
  \centering
  \resizebox{1\textwidth}{!}{\includegraphics{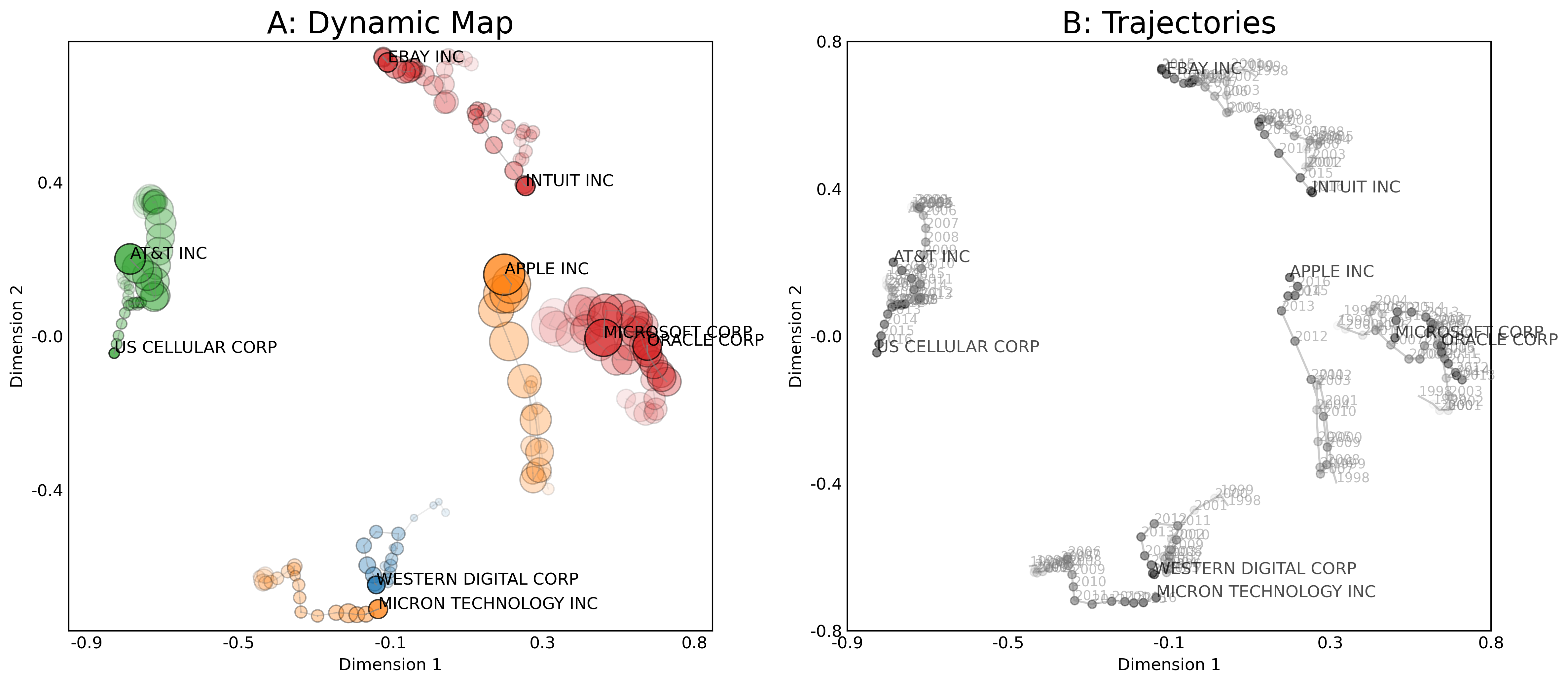}}
  \caption{\label{fig:draw-dynamic-map-and-trajectories} Dynamic maps for TNIC sample}
\end{figure}

The resultant map is shown in the left graph of Figure \ref{fig:draw-dynamic-map-and-trajectories}.

Alternatively, to focus solely on the movement paths, \code{draw_trajectories()} 
offers a specialized view highlighting the trajectory of each object through time. The result can be seen in the 
right part of Figure \ref{fig:draw-dynamic-map-and-trajectories}.

\begin{CodeChunk}
\begin{CodeInput}
>>> from evomap.printer import draw_trajectories
>>> draw_trajectories(X_t,
...                   labels=labels,
...                   period_labels=periods,
...                   show_axes=True,
...                   ax=ax[1])
\end{CodeInput}
\end{CodeChunk}

\subsection{Evaluation}

While visual exploration allows users to examine the mapping results, 
it is essential to evaluate how well they represent the underlying data. This can be achieved through:

\vbox{
  \begin{enumerate}
    \item \textbf{The cost function}, which is useful for comparing solution quality at different points in time % 
    or comparing different settings within the same method.  
    \item \textbf{Dedicated metrics}, offered in the \code{metrics} module, which provide a systematic way to %
    assess different aspects of mapping quality and/or compare it across methods. 
  \end{enumerate}
}

In the TNIC dataset example, we can assess how EvoMap's regularization affects individual 
mapping quality by comparing the average Stress against a sequence generated independently 
(i.e., with alpha set to 0). The increase in Stress is negligible (+0.0066), which indicates a minimal decrease in static quality 
for the benefit of alignment.

\begin{CodeChunk}
\begin{CodeInput}
>>> cmds_indep = [CMDS().fit_transform(D) for D in D_t]
>>> mds_indep = EvoMDS(
...     alpha=0,
...     init=cmds_indep,
...     mds_type='ordinal')
>>> X_t_indep = mds_indep.fit_transform(D_t)
>>> print(mds_indep.cost_static_avg_.round(4))
>>> print(evomds.cost_static_avg_.round(4))
\end{CodeInput}
\begin{CodeOutput}
0.1875
0.1941
\end{CodeOutput}
\end{CodeChunk}

Table \ref{tab:evaluation-metrics} presents the additional evaluation metrics available in the \code{metrics} module. 
The table lists the metric, its purpose, application scope (single snapshot vs. sequence), and interpretation range. We refer to 
\cite{Matthe+Ringel+Skiera:2023} for more
details on these metrics and their underlying intuition.

\begin{table}[ht]
  \centering
  \begin{tabular}{p{3cm}p{4.5cm}p{2.5cm}p{3cm}}
  \hline
  \textbf{Metric} & \textbf{Intuition} & \textbf{Snapshot \newline vs. sequence} & \textbf{Range} \\ \hline
  Misalignment & Total length of all movement paths. & Sequence & 0 (good) to Inf (poor) \\ 
  Alignment & Average cosine similarity of subsequent positions. & Sequence & -1 (poor) to 1 (good) \\ 
  Hit-Rate & Agreement in nearest neighbors between data and the map. & Snapshot & 0 (poor) to 1 (good) \\ 
  Adjusted \newline Hit-Rate & Hit-Rate, adjusted for random agreement. & Snapshot & 0 (poor) to 1 (good) \\ 
  Average Hit-Rate & Average Hit-Rate over multiple periods. & Sequence & 0 (poor) to 1 (good) \\ 
  Average Adjusted Hit-Rate & Average adjusted Hit-Rate over multiple periods. & Sequence & 0 (poor) to 1 (good) \\ 
  Persistence & Correlation of subsequent movement vectors on the map. & Sequence & -1 (poor) to 1 (good) \\ \hline
  \end{tabular}
  \caption{\label{tab:evaluation-metrics} Evaluation metrics}
\end{table}

In the TNIC example, we compute the following metrics: alignment and persistence. The former measures 
how well subsequent positions align with each other, while the latter assesses the smoothness of movement paths. 
We compute each metric for EvoMap's output and two benchmarks: independent application of MDS and ex-post alignment of independent mapping via 
Procrustes Analysis. The required transformations (rotations/reflections) are available through the \code{align_maps()} function,
while the evaluation metrics can be imported from the \code{evomap.metrics} module:

\begin{CodeChunk}
\begin{CodeInput}
>>> from evomap.transform import align_maps
>>> X_t_indep_aligned = align_maps(X_t_indep, X_t_indep[0])

>>> from evomap.metrics import misalign_score, persistence_score
>>> misalignments = [
...     misalign_score(X) 
...     for X in [X_t_indep, X_t_indep_aligned, X_t]]
>>> persistences = [
...     persistence_score(X) 
...     for X in [X_t_indep, X_t_indep_aligned, X_t]]
>>> avg_stresses = [
...    evomds_indep.cost_static_avg_,
...    evomds_indep.cost_static_avg_, 
...    evomds.cost_static_avg_]
>>> df_eval = pd.DataFrame({
...     'misalign_score': misalignments,
...     'persistence_score': persistences,
...     'average_stress': avg_stresses
... }, index=['Independent MDS', 'Independent MDS + Alignment', 'EvoMDS'])
>>> print(df_eval.round(4))
\end{CodeInput}
\end{CodeChunk}

Table \ref{tab:evaluation-results} displays the results. EvoMap's solution is 
better aligned and movement paths are smoother (thus, easier to interpret), while static mapping quality is not 
substantially reduced. 

\begin{table}[ht]
  \centering
  \begin{tabular}{lp{2.5cm}p{2cm}p{2cm}p{2cm}}
    \hline
    Test & Misalignment Score & Persistence Score & Static Cost \\
    \hline
    Independent MDS & 0.9737 & -0.5589 & 0.1875 \\
    Independent MDS \newline + Alignment & 0.3068 & -0.2697 & 0.1875 \\
    EvoMDS & 0.0352 & 0.7433 & 0.1941 \\
    \hline
  \end{tabular}  
  \caption{\label{tab:evaluation-results} Evaluation results}
\end{table}

\FloatBarrier

\begin{figure}[hbt!]
  \centering
  \resizebox{1\textwidth}{!}{\includegraphics{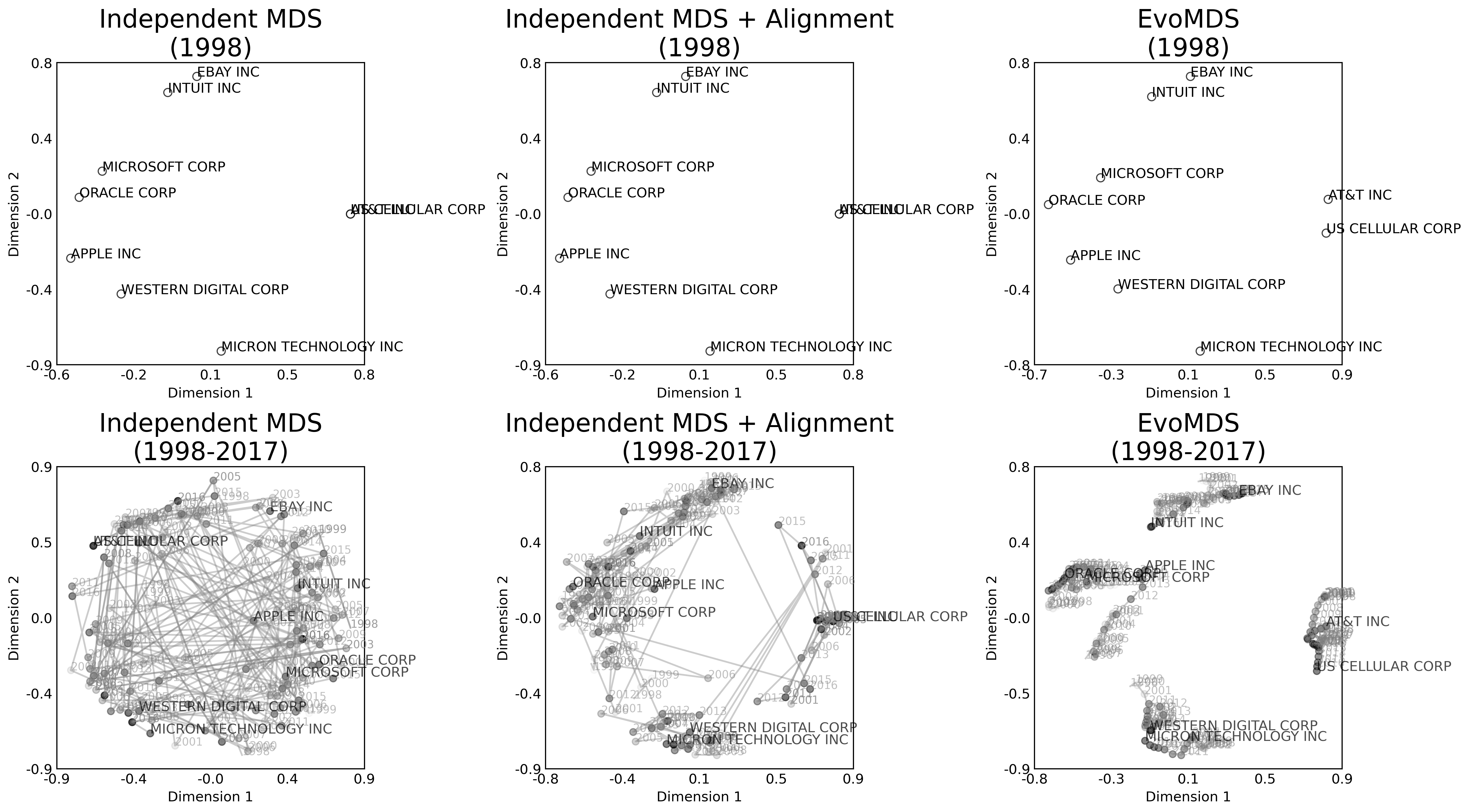}}
  \caption{\label{fig:visual-comparioson} Visual comparison of EvoMap, independent MDS, and aligned MDS}
\end{figure}

These quantitative differences closely correspond to visual differences in these configurations, which 
are illustrated in Figure \ref{fig:visual-comparioson}. To make all three approaches easy to compare visually, 
we also apply a fixed linear transformation to the sequence of configurations generated by EvoMap, such that 
the coordinate systems of all three solutions align. Figure \ref{fig:visual-comparioson} displays a static view of the first 
configuration in each sequence (top row) as well as a dynamic view of the subsequent movement paths (bottom row):

\begin{CodeChunk}
\begin{CodeInput}
>>> fig, ax = plt.subplots(2, 3, figsize=(18, 10))
>>> X_t_aligned = align_maps(X_t, X_t_indep[0], 'fixed')
>>> draw_map(X_t_indep[0],
...          label=labels,
...          ax=ax[0, 0])
>>> draw_map(X_t_indep_aligned[0],
...          label=labels,
...          ax=ax[0, 1])
>>> draw_map(X_t_aligned[0],
...          label=labels,
...          ax=ax[0, 2])
>>> draw_trajectories(X_t_indep,
...                   labels=labels,
...                   period_labels=periods,
...                   show_axes=True,
...                   ax=ax[1, 0])
>>> draw_trajectories(X_t_indep_aligned,
...                   labels=labels,
...                   period_labels=periods,
...                   show_axes=True,
...                   ax=ax[1, 1])
>>> draw_trajectories(X_t_aligned,
...                   labels=labels,
...                   period_labels=periods,
...                   show_axes=True,
...                   ax=ax[1, 2])
>>> titles = [
...     'Independent MDS\n(1998)',
...     'Independent MDS + Alignment\n(1998)',
...     'EvoMDS\n(1998)',
...     'Independent MDS\n(1998-2017)',
...     'Independent MDS + Alignment\n(1998-2017)',
...     'EvoMDS\n(1998-2017)'
... ]
>>> for i in range(2):
...     for j in range(3):
...         ax[i, j].set_title(titles[i * 3 + j])
...         ax[i, j].set_xlabel('Dimension 1')
...         ax[i, j].set_ylabel('Dimension 2')
\end{CodeInput}
\end{CodeChunk}

As seen in the first row of Figure \ref{fig:visual-comparioson}, the static configurations produced by EvoMap 
are indeed almost identical to those produced by independent MDS, in line with the minimal increase in average Stress 
shown in Table \ref{tab:evaluation-metrics}. As seen in the second row, the temporal evolution of the configurations,
however, is much smoother for EvoMap, with the movement paths being more aligned and less zigzagged. 

\section{Advanced considerations}

To close the presentation of the \pkg{evomap} package, this section expands on two considerations that are 
relevant in many applications: finding good hyperparameter values and  dealing with unbalanced data.

\subsection{Finding good hyperparameter values} \label{sec:hyperparameters}

Determining good values for the hyperparameters \(\alpha\) and \(p\) is a key step in using EvoMap  
effectively. Naturally, there is no one-size-fits-all solution, as the appropriateness of these values  
varies significantly across different applications. For instance, this choice can be affected by the scale  
of the static cost function, the number of objects, and the length of the observation period. Consequently,  
identifying suitable hyperparameter values typically involves an exploratory process  
tailored to each application.

To support this process, the package includes two additional functionalities:
\begin{enumerate}
  \item \textbf{Grid search}, which explores a user-defined grid of \(\alpha\) and \(p\)  
  combinations and helps visualize trade-offs between competing objectives (e.g., goodness of fit  
  vs. temporal alignment) and identify promising regions of the parameter space for further inspection.

  \item \textbf{Bayesian optimization}, which searches for hyperparameter values that minimize 
  a desired combination of evaluation metrics. Technically, it treats hyperparameter search as a multi-objective 
  optimization problem with a scalarized additive objective function (the weighted sum 
  of different evaluation metrics), and iteratively proposes candidate values to find a combination of 
  hyperparameter values that minimizes this objective.
\end{enumerate}

In short, the grid search is well suited for initial exploration and visual analysis of trade-offs,  
while Bayesian optimization provides a more principled and automated way to identify appropriate  
parameter combinations within a limited number of evaluations. 

To run the \textbf{grid search}, the user must first define a hyperparameter grid, select a set of evaluation metrics, 
 and call the \code{grid\_search()} function. 

\vbox{
\begin{CodeChunk}
\begin{CodeInput}
>>> param_grid = {
...     'alpha': np.linspace(0, 1.5, 15), 
...     'p': [1, 2]}
>>> from evomap.metrics import avg_hitrate_score
>>> metrics = [misalign_score, persistence_score, avg_hitrate_score]
>>> metric_labels = ['Misalignment', 'Persistence', 'Hitrate']

>>> model = EvoMDS(init=cmds_t, mds_type='ordinal')
>>> df_grid_results = model.grid_search(
...     Xs=D_t,
...     param_grid=param_grid,
...     eval_functions=metrics,
...     eval_labels=metric_labels).reset_index()
\end{CodeInput}
\end{CodeChunk}
}

The results can then be visualized to explore attractive regions in the hyperparameter space.

\begin{CodeChunk}
\begin{CodeInput}
>>> from matplotlib import colormaps
>>> fig, ax = plt.subplots(1, 3, figsize=(18, 5))
>>> ps = sorted(df_grid_results['p'].unique())
>>> colors = colormaps['Set1'].resampled(len(ps))
>>> metrics = ['Misalignment', 'Persistence', 'cost_static_avg']
>>> titles = ['Misalignment', 'Persistence', 'Avg. Static Cost']
>>> for i, (metric, title) in enumerate(zip(metrics, titles)):
...     for j, p in enumerate(ps):
...         subset = df_grid_results[df_grid_results['p'] == p]
...         color = colors(j)
...         ax[i].plot(
...             subset['alpha'],
...             subset[metric],
...             label=f'p = {p}' if i == 0 else None,
...             color=color,
...             alpha=0.5
...         )
...         ax[i].scatter(
...             subset['alpha'],
...             subset[metric],
...             color=color,
...             s=50,
...             edgecolor='black',
...             linewidth=0.6
...         )
...     ax[i].set_title(title)
...     ax[i].set_xlabel('Alpha')
...     ax[i].set_ylabel('Score')
...     ax[i].grid(True, linestyle=':', alpha=0.6)
...     ax[i].tick_params(labelsize=10)
>>> ax[0].legend(title='p', fontsize=10, title_fontsize=11)
\end{CodeInput}
\end{CodeChunk}

\begin{figure}[hbt!]
  \centering
  \resizebox{\textwidth}{!}{\includegraphics{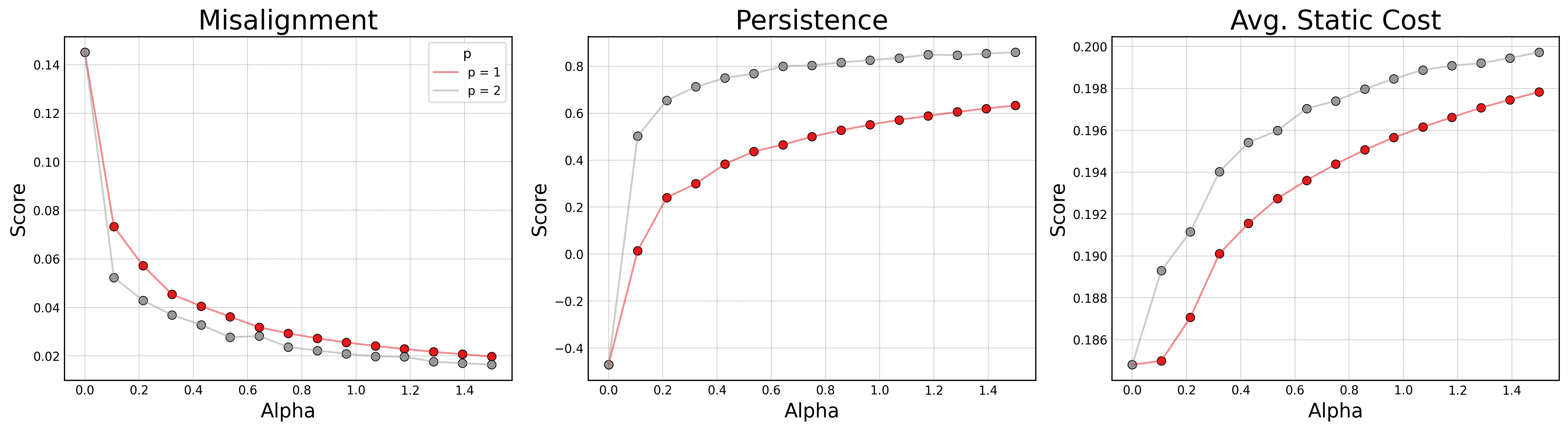}}
  \caption{\label{fig:grid-search} Grid search results}
\end{figure}

Here, the results in Figure \ref{fig:grid-search} reveal a sharp initial decrease in Misalignment scores 
for $\alpha$ values lower than .4. As improvements in Misalignment and Persistence scores quickly show decreasing
marginal returns beyond this value, while static cost--that is, badness-of-fit--continues to increase, this suggests
$\alpha$ values below .4 a viable range for further examination. 
Within this range, setting $p = 2$ enhances Persistence without substantially compromising static fit, 
as indicated by the only moderate increase in average Stress.

While straightforward, the grid search has clear disadvantages. The evaluation metrics are not necessarily smooth
in the hyperparameter space, which might demand a large number of evaluations to identify suitable values.
This can quickly become computationally expensive and requires a possibly large number of evaluations, which is time consuming--especially for large datasets.

A more targeted and efficient strategy is to use \textbf{Bayesian optimization}. Bayesian optimization has been shown to 
perform well for hyperparameter tuning in static mapping  contexts, most notably in the STOPS framework \citep{Rusch+Mair+Hornik:2022}, 
where it is used to optimize parameters based on structural properties of the output configurations (referred to as  
\emph{c-structuredness} indices). The \pkg{evomap} package builds on this idea by applying Bayesian  
optimization to dynamic mapping, where it is used to optimize EvoMap's hyperparameters based on desirable  
properties of the resulting configuration sequences—such as goodness-of-fit alongside temporal alignment   
and smoothness.

Within the \pkg{evomap} package, the \code{bayesian_search()} function implements Bayesian optimization to find suitable hyperparameter values.
Here, the user must first define a search space for $alpha$ and $p$, specify the evaluation functions, and then call the \code{bayesian\_search()} method.

When calling this function, the user also specifies weights which determine the relative importance of each evaluation function. 
The scalarized objective function is \emph{additive}, that is, a weighted sum of the provided 
evaluation functions, where the first weight corresponds to the static cost function of the mapping method used (here: Stress),
while the remaining weights correspond to the evaluation functions the user provided.

In the example below, we provide the \code{misalign\_score} and an inverted \code{persistence\_score} as evaluation functions. 
We compute and use the inverted persistence score as $1 - \text{persistence}$ to ensure all metrics are 
\emph{minimization-aligned} (i.e., lower values indicate better solutions). 
We then assign the weights \code{[0.95, 0.03, 0.02]}, which corresponds to optimizing an objective function composed of 
$0.95 \times \text{Stress} + 0.03 \times \text{Misalignment} + 0.02 \times (1 - \text{Persistence})$. 
Put differently, the combined objective function gives 95\% weight to the static cost function, 
3\% to the misalignment score, and 2\% to the inverted persistence score.

\vbox{
\begin{CodeChunk}
\begin{CodeInput}
>>> from skopt.space import Real, Integer
>>> search_space = [
...     Real(0.0, 1.0, name='alpha'),
...     Integer(1, 2, name='p')]

>>> def inverted_persistence(X_t):
...     return 1 - persistence_score(X_t)

>>> model = EvoMDS(init=cmds_t, mds_type='ordinal')
>>> results = model.bayesian_search(
...     Xs=D_t,
...     search_space=search_space,
...     eval_functions=[misalign_score, inverted_persistence],
...     eval_labels=['misalign', 'persistence_inverted'],
...     weights=[0.95, 0.03, 0.02],
...     opt_params={'n_calls': 20, 'n_initial_points': 3,
...                 'acq_func': 'EI', 'random_state': 42}
... )
\end{CodeInput}
\begin{CodeOutput}
Best result found:  
alpha: 0.1442
p: 2
cost_static_avg: 0.1905
misalign: 0.0471
persistence_inverted: 0.4300
combined_loss: 0.1910
\end{CodeOutput}
\end{CodeChunk}
}

Again, the results can be visualized to explore the cost surface across the hyperparameter space. 

\begin{CodeChunk}
\begin{CodeInput}
>>> fig, ax = plt.subplots(1, 4, figsize=(24, 5))
>>> metrics = [
...     'cost_static_avg',
...     'misalign',
...     'persistence_inverted',
...     'combined_loss']
>>> titles = [
...     'Avg. Static Cost',
...     'Misalignment',
...     '1 - Persistence',
...     'Combined Loss']
>>> ps = sorted(results['p'].unique())
>>> colors = colormaps['Set1'].resampled(len(ps))
>>> for i, (metric, title) in enumerate(zip(metrics, titles)):
...     for j, p in enumerate(ps):
...         df_p = results[results['p'] == p].sort_values('alpha')
...         ax[i].plot(df_p['alpha'], df_p[metric], color=colors(j), alpha=0.5)
...         ax[i].scatter(df_p['alpha'], df_p[metric], color=colors(j), s=50,
...                       edgecolor='black', linewidth=0.6,
...                       label=f'p = {p}' if i == 0 else None)
...     ax[i].set(title=title, xlabel='Alpha', ylabel='Score')
...     ax[i].grid(True, linestyle=':', alpha=0.6)
...     ax[i].tick_params(labelsize=10)
>>> ax[0].legend(title='p', fontsize=10, title_fontsize=11)
\end{CodeInput}
\end{CodeChunk}

\begin{figure}[hbt!]
  \centering
  \resizebox{\textwidth}{!}{\includegraphics{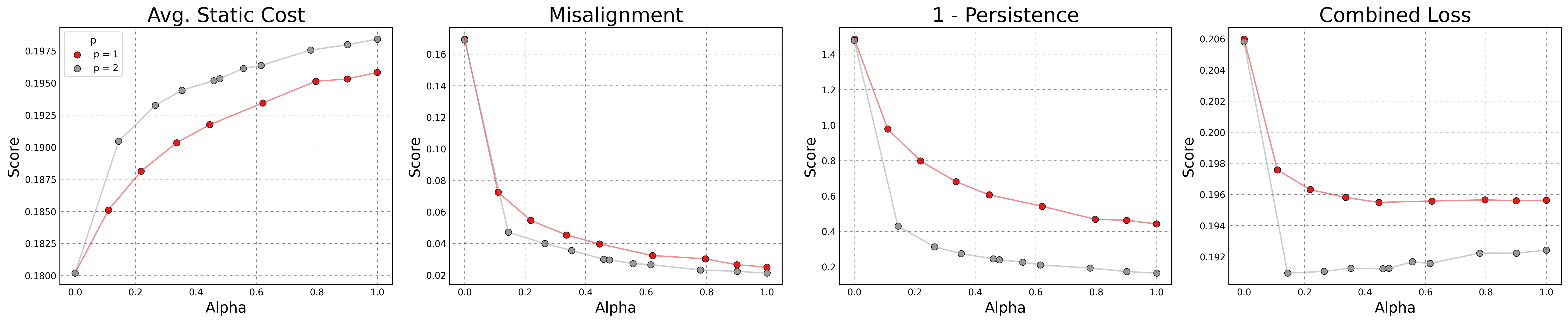}}
  \caption{\label{fig:bayesian-search} Bayesian search results}
\end{figure}

The results indicate that a hyperparameter combination of $\alpha \approx 0.14$ and $p = 2$ minimizes 
the estimated combined loss function of static cost (Stress), misalignment, and persistence 
(see Figure~\ref{fig:bayesian-search}).\footnote{Results from Bayesian optimization may vary slightly across 
platforms (e.g., Windows, macOS, Linux), due to platform-specific differences in floating-point arithmetic, 
linear algebra backends (e.g., BLAS/LAPACK), or internal solver tolerances. These differences are typically 
negligible and do not meaningfully affect the resulting configuration.}

Note that while the presented example used metrics related to the temporal evolution of the resulting configurations, 
the Bayesian optimization framework can, in principle, incorporate any metric—including additional structural properties, 
as demonstrated in the STOPS framework by \cite{Rusch+Mair+Hornik:2022}—as well as tune other method-specific hyperparameters.

A general question concerns the appropriate trade-off between goodness-of-fit and temporal properties of the configurations. 
While no universal guideline exists, a 5\% trade-off in fit, as in the example above, may represent a good starting point for MDS-based methods. 
If large trade-offs are required to produce aligned and smooth trajectories, this may suggest 
that the underlying evolution is genuinely erratic rather than gradual. It is important to remember that 
mapping methods---including the dynamic variants implemented in this package---are primarily exploratory tools. 
As such, users are encouraged to experiment with different hyperparameters, inspect results visually, 
and assess the robustness of the resultant configurations to these choices.

\subsection{Dealing with unbalanced data} \label{sec:unbalanced_data}

Dynamic mapping often encounters scenarios where the set of objects changes over time,  
such as firms entering or exiting a market. The \pkg{evomap} package accommodates such unbalanced data  
through inclusion vectors.

To maintain a consistent data format across time, the package requires each relationship matrix  
in the sequence to have the same dimensions---that is, each matrix must include the same set  
of objects in the same order, even if some objects are not observed at certain time points.  
This is achieved by:

\begin{enumerate}
  \item \textbf{Balancing the data}, which entails constructing a sequence of square matrices with consistent dimensions  
  and object ordering which cover all objects ever observed during the analysis period.  
  For time points where an object is not observed, one can fill the corresponding row and column  
  with an arbitrary placeholder (e.g., 0).
  \item \textbf{Creating inclusion vectors}, which entails generating a binary vector, 
  indicating, for each time point, whether each object is observed (1) or not (0). These vectors determine which  
  entries are considered during mapping.
\end{enumerate}

The \code{preprocessing} module's \code{expand_matrices()} function simplifies this process by automatically balancing the data and generating inclusion vectors.

Consider an extended TNIC dataset including Netflix, which only appears in the dataset after its IPO in 2002, 
resulting in unbalanced data. The following steps illustrate how to prepare and analyze this dataset:

\textbf{1. Load and balance the data:}

\begin{CodeChunk}
\begin{CodeInput}
>>> from evomap.preprocessing import expand_matrices
>>> data_unbalanced = load_tnic_sample_tech(unbalanced=True)
>>> S_t, labels = edgelist2matrices(
...     data_unbalanced,
...     score_var='score',
...     id_var_i='name1',
...     id_var_j='name2',
...     time_var='year')
>>> print(S_t[0].shape)   # Output: (9, 9)
>>> S_t, inc_t, labels = expand_matrices(S_t, labels)
>>> print(S_t[0].shape)   # Output: (10, 10)
>>> print(inc_t[0])       # Output: [1 1 1 1 1 1 1 1 1 0]
\end{CodeInput}
\end{CodeChunk}

\textbf{2. Transform data and fit EvoMap:}

\begin{CodeChunk}
\begin{CodeInput}
>>> D_t = [sim2diss(S, transformation='mirror') for S in S_t]
>>> init_t = [np.concatenate([cmds, np.array([[0, 0]])], axis=0)
...           for cmds in cmds_t]
>>> evomds_unbalanced = EvoMDS(
...     alpha=0.75,
...     p=2,
...     init=init_t,
...     mds_type='ordinal')
>>> X_t = evomds_unbalanced.fit_transform(D_t, inclusions=inc_t)
\end{CodeInput}
\end{CodeChunk}
  
\textbf{3. Visualize the results:}

\begin{CodeChunk}
\begin{CodeInput}
>>> fig, ax = plt.subplots(1, 2, figsize=(14, 6))
>>> draw_map(X_t[0],
...          inclusions=inc_t[0],
...          label=labels,
...          show_axes=True,
...          ax=ax[0])
>>> draw_map(X_t[-1],
...          inclusions=inc_t[-1],
...          label=labels,
...          show_axes=True,
...          ax=ax[1])
\end{CodeInput}
\end{CodeChunk}
  
\begin{figure}[hbt!]
  \centering
  \resizebox{\textwidth}{!}{\includegraphics{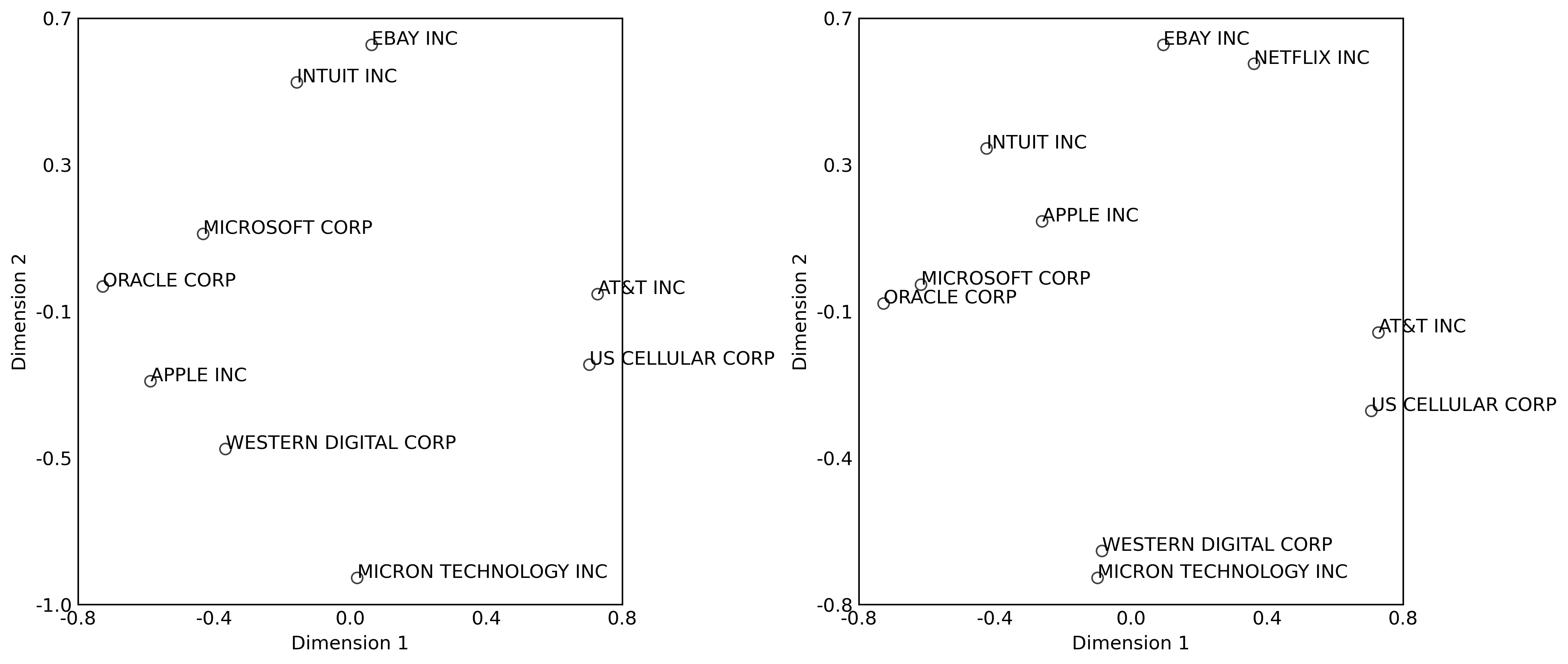}}
  \caption{\label{fig:unbalanced-data} EvoMap snapshots for 1998 and 2017, using unbalanced data}
\end{figure}

Figure \ref{fig:unbalanced-data} shows the first and last snapshot of the generated sequence. As seen in the Figure,
EvoMap has successfully mapped the unbalanced data. The inclusion vectors ensure that EvoMap accurately reflects the 
entry and exit of firms: Netflix is not present on the left map corresponding to 1998, while it appears close to 
software-focused firms (eBay) on the right map corresponding to 2017.

\section{Performance on simulated data} \label{sec:Simulations}

\subsection{Recovery of simulated movement paths}

A natural limitation of empirical data---such as the example shown in the previous section---is that the 
underlying \emph{ground truth} is typically unknown. In our context, this refers to the true positions of 
objects in the latent space and how they evolve over time. Methods like MDS are known 
to be sensitive to even small changes in their input data, which means that measurement noise could quickly 
distort the resulting configuration, which can impair their ability to accurately recover the true positions from
the provided data.

To illustrate how EvoMap performs in a setting where the ground truth is known, we report the results of 
an additional simulation study. Here, we simulate a sequence of known positions and measure a sequence of dissimilarity matrices based on 
these positions. This allows us to directly evaluate how accurately EvoMap can recover the simulated positions. 
For a more elaborate and extensive simulation study, we refer the reader to \cite{Matthe+Ringel+Skiera:2023}.

In our simulation, we generate the movement of $n = 6$ objects in a two-dimensional space over 
$t = 10$ time periods. The trajectories follow a momentum-based random walk: each object's position at 
time $t$ is determined by its prior velocity and a random perturbation. This creates movement paths that are 
random yet smooth and consistent across time. These simulated paths serve as the known ground truth 
against which EvoMap's output can be compared.

We start by simulating the movement paths of the objects, which are represented as two-dimensional coordinates
in a latent space.

\begin{CodeChunk}
\begin{CodeInput}
>>> np.random.seed(123)
>>> def simulate_paths(n=6, t=10, scale=1, noise=0.25, momentum=0.6):
...     X, dX = np.random.randn(n, 2) * scale, np.random.randn(n, 2) * noise
...     out = [X.copy()]
...     for _ in range(1, t):
...         dX = momentum * dX + (1 - momentum) * np.random.randn(n, 2) * noise
...         out.append((X := X + dX).copy())
...     return out
>>> X_true = simulate_paths()
\end{CodeInput}
\end{CodeChunk}

For each time step, we compute pairwise Euclidean distances between all objects--possibly subject to small amounts of 
 noise--and use these as input to EvoMap. 

\begin{CodeChunk}
\begin{CodeInput}
>>> from scipy.spatial.distance import pdist, squareform
>>> def measure_distances(X_true, noise=0):
...     X_noisy = [X + np.random.randn(*X.shape) * noise for X in X_true]
...     D_sim = [squareform(pdist(X)) for X in X_noisy]
...     return D_sim
>>> D_sim = measure_distances(X_true, noise=0)
>>> evomds_model = EvoMDS(mds_type='ratio', alpha=0.2, p=1, n_inits=10)
>>> X_evo = evomds_model.fit_transform(D_sim)
>>> X_evo_aligned = align_maps(X_evo, X_true[0], 'fixed')
\end{CodeInput}
\end{CodeChunk}

To evaluate recovery, we assess the average Procrustes distance across all time periods (that is, the Euclidean Distance
between the recovered and simulated configurations, after applying Procrustes Analysis for optimal scaling/rotation/reflection). 

\begin{CodeChunk}
\begin{CodeInput}
>>> from scipy.spatial import procrustes
>>> print(f'Stress EvoMDS: {evomds_model.cost_static_avg_:.4f}')
>>> distances = [procrustes(X_true[t], X_evo[t])[2] for t in range(len(X_true))]
>>> mean_distance = np.mean(distances)
>>> print(f'Procrustes Distance: {mean_distance:.4f}')
>>> fig, ax = plt.subplots(1, 2, figsize=(10, 5))
>>> draw_trajectories(
...     X_true,
...     labels=[str(i) for i in range(X_true[0].shape[0])],
...     title='A: Simulated',
...     ax=ax[0])
>>> draw_trajectories(
...     X_evo_aligned,
...     labels=[str(i) for i in range(X_true[0].shape[0])],
...     title='B: EvoMDS',
...     ax=ax[1])
\end{CodeInput}
\begin{CodeOutput}
Stress EvoMDS: 0.0111
Procrustes Distance: 0.0003
\end{CodeOutput}
\end{CodeChunk}

\begin{figure}[hbt!]
  \centering
  \resizebox{1\textwidth}{!}{\includegraphics{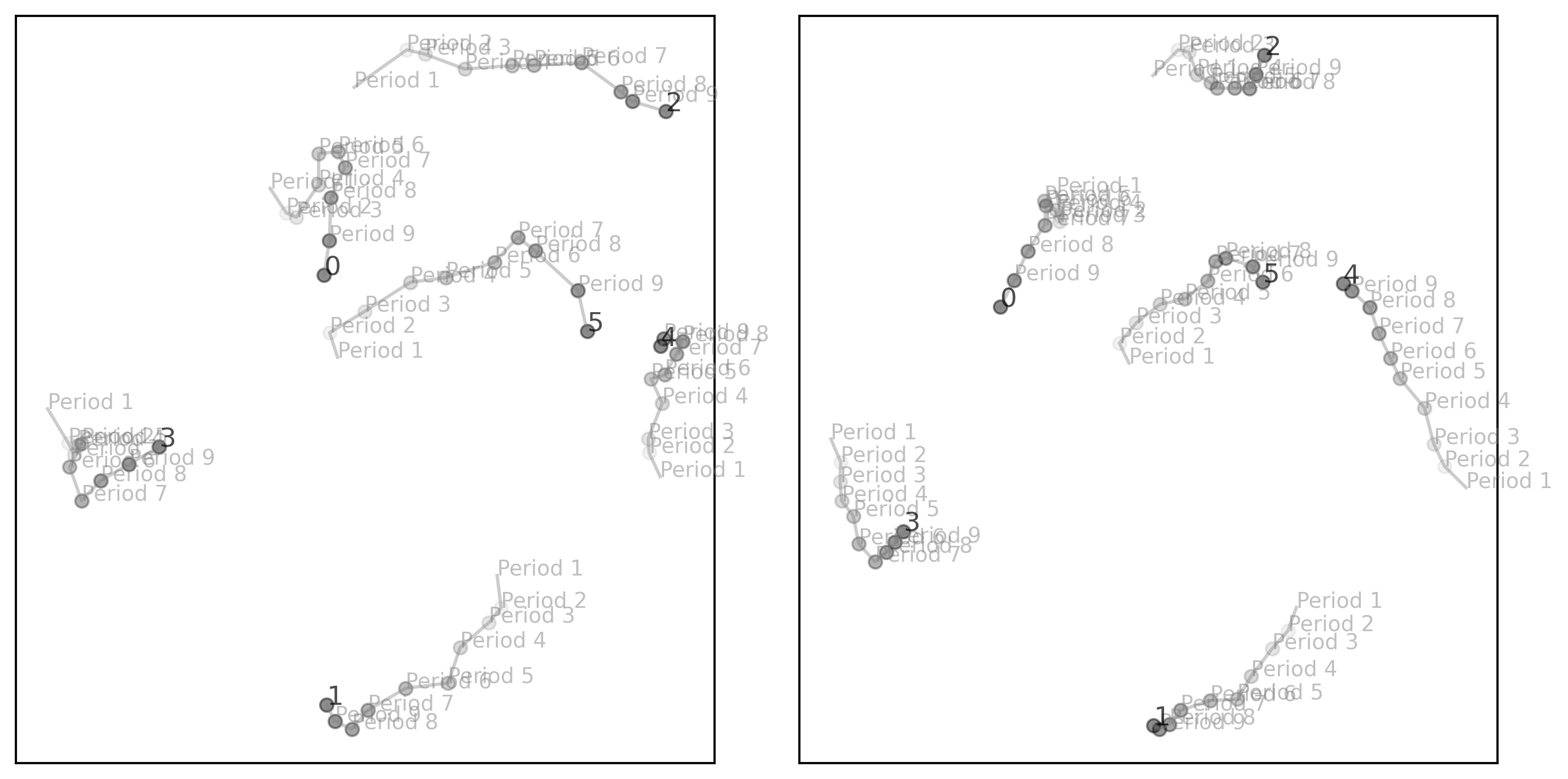}}
  \caption{\label{fig:simulated-vs-recovered} Visual comparison of simulated vs. recovered movement paths}
\end{figure}

Quantitatively, both the average Procrustes distance and the average Stress of EvoMap are very low, indicating that EvoMap
effectively fits the measured distances and recovers the simulated positions well. Visually, Figure \ref{fig:simulated-vs-recovered}
confirms that the simulated and recovered movement paths are nearly identical (after scaling, rotating, and reflecting them 
such that their coordinate systems align).

\FloatBarrier

Next, we repeat the same procedure, but add varying degrees of measurement noise after measuring distances among the 
simulated positions. 

\begin{CodeChunk}
\begin{CodeInput}
>>> results = []
>>> for noise in [0.01, 0.5]:
...     for i in range(10):
...         X_true = simulate_paths()
...         D_sim = measure_distances(X_true, noise=noise)
...         for alpha in [0, 0.3]:
...             evomds_model = EvoMDS(
...                 mds_type='ratio',
...                 alpha=alpha,
...                 n_inits=10)
...             X_hat = evomds_model.fit_transform(D_sim)
...             stress = evomds_model.cost_static_avg_
...             X_hat_aligned = align_maps(X_hat, X_true[0])
...             proc_dist = np.mean([
...                 procrustes(X_true[t], X_hat_aligned[t])[2]
...                 for t in range(len(X_true))])
...             metrics = {
...                 'noise': noise,
...                 'iteration': i,
...                 'alpha': alpha,
...                 'stress': stress,
...                 'procrustes': proc_dist}
...             results.append(metrics)
>>> df_results = pd.DataFrame(results)
>>> df_summary = (
...     df_results
...     .groupby(['noise', 'alpha'])[['stress', 'procrustes']]
...     .mean(numeric_only=True)
...     .round(4)
... )
>>> print(df_summary)
\end{CodeInput}
\end{CodeChunk}

Results are summarized in Table~\ref{tab:simulation-results}. As the regularization parameter 
$\alpha$ increases, the overall stress values also increase. In the low-noise condition, this 
corresponds to a slight, almost negligible, decrease in the Procrustes distance between the recovered 
and true positions. That is, stronger temporal regularization does not impair recovery performance 
when the input data are subject to only low levels of noise. It neither substantially helps nor hurts the recovery 
itself—beyond the improved temporal alignment effects discussed earlier.

In the noisy condition, a more interesting pattern emerges. While increasing $\alpha$ results in 
substantially higher stress values—reflecting a reduced fit to the noisy dissimilarity matrices—it 
also substantially reduces the Procrustes distance to the true positions. This indicates that, 
despite fitting the observed data less closely, EvoMap recovers the underlying ground truth 
\emph{more} accurately when regularization is applied.

These findings highlight a key advantage of EvoMap: its temporal regularization may lead to 
a slightly worse fit to the observed (and potentially noisy) input data, but it can substantially 
improve recovery of the true latent structure—at least under conditions similar to the simulated setup. 
Thus, when the observed dissimilarities represent noisy measurements of an underlying smooth process, 
EvoMap helps filter out the noise and recover a more faithful trajectory in the lower-dimensional space.

\begin{table}[hbt!]
  \centering
  \small
  \renewcommand{\arraystretch}{1.1}
  \begin{tabular}{p{1.2cm}p{1.2cm}p{3.2cm}p{4cm}}
  \hline
  Noise & Alpha & Avg. Stress & Avg. Procrustes Distance \\
  \hline
  0.01 & 0.0 & 0.0036 & 0.0067 \\
       & 0.3 & 0.0189 & 0.0016 \\
  0.50 & 0.0 & 0.0015 & 0.1262 \\
       & 0.3 & 0.1027 & 0.0719 \\
  \hline
  \end{tabular}
  \caption{\label{tab:simulation-results} Recovery accuracy under varying noise levels and alpha values}
\end{table}

While this simulation mostly serves as a small illustration, we refer to a much more elaborate simulation study
(including more complex data structure, movement patterns, and simulation parameters) in \cite{Matthe+Ringel+Skiera:2023}. 

\subsection{Complexity and runtime} \label{sec:Complexity}

Finally, we briefly illustrate the runtime characteristics of the current implementation. 
Most of the computational complexity arises from the static mapping method employed at each time step. 
The cost of these methods scales non-linearly with the number of objects, \(n\), which makes it the primary 
bottleneck.

The exact complexity depends on the specific mapping method used within EvoMap. For instance, classical MDS has complexity of up to
\(\mathcal{O}(n^3)\), depending on the specific variant used. EvoMap evaluates the static cost function at each time point,
making it linear in the number of time points, thus \(\mathcal{O}(n^3T)\). 

By contrast, the temporal cost component of EvoMap adds computational costs that are only linear in the number of 
objects, as the k-backward differences can be computed independently for each object. It also scales linearly in the 
number of periods, which determines how many differences need to be computed per object, and in $p$, which determines the maximum number 
of differences computed per object and time point. As $n>>p$, the complexity of the total cost function 
is therefore dominated by the static component, and thus \(\mathcal{O}(n^3 T)\) or better. 

To provide practical guidance on what to expect in terms of runtime, this section presents a simple 
benchmark study using simulated datasets of varying sizes. 

\begin{CodeChunk}
\begin{CodeInput}
>>> import time
>>> from evomap.mapping import MDS
>>> def benchmark_evomds_vs_static_mds(
...     n_list=[10, 50, 100],
...     t_list=[10, 50, 100]
... ):
...     times_evomds = []
...     times_mds = []
...     for n in n_list:
...         for t_periods in t_list:
...             X_true = simulate_paths(n, t_periods)
...             D_sim = measure_distances(X_true)
...             model = EvoMDS(
...                 alpha=0.3,
...                 mds_type='ratio',
...                 tol=0,
...                 n_iter=750)
...             start = time.time()
...             model.fit_transform(D_sim)
...             elapsed_evo = time.time() - start
...             times_evomds.append({
...                 'n': n, 't': t_periods, 'seconds': elapsed_evo})
...             start = time.time()
...             for D_t in D_sim:
...                 mds = MDS(mds_type='ratio', tol=0, n_iter=750)
...                 mds.fit(D_t)
...             elapsed_mds = time.time() - start
...             times_mds.append({
...                 'n': n, 't': t_periods, 'seconds': elapsed_mds})
...     df_evomds = (
...         pd.DataFrame(times_evomds)
...           .pivot(index='n', columns='t', values='seconds')
...           .round(2))
...     df_mds = (
...         pd.DataFrame(times_mds)
...           .pivot(index='n', columns='t', values='seconds')
...           .round(2))
...     return df_evomds, df_mds
>>> df_evo, df_mds = benchmark_evomds_vs_static_mds()
>>> print("EvoMDS Benchmarking Results:")
>>> print(df_evo)
>>> print("Static MDS Benchmarking Results:")
>>> print(df_mds)
\end{CodeInput}
\begin{CodeOutput}
EvoMDS Benchmarking Results:
t      10     50      100
n                        
10   11.80  30.00   54.11
50   13.14  33.42   66.04
100  17.90  47.17  119.00

Static MDS Benchmarking Results:
t      10     50      100
n                        
10    6.03  39.72   78.74
50    9.97  44.91   83.24
100  11.45  60.75  117.59
\end{CodeOutput}
\end{CodeChunk}

While the exact runtimes naturally depend on the hardware and software environment, these results 
demonstrate that \pkg{evomap} can handle relatively large datasets with a moderate number of objects 
and time periods. On a current MacBook Pro with an M2 Max chip, running macOS 15.1.1 and Python 3.13.2, 
a single run of the most demanding case (100 objects over 100 time points) completes in roughly two minutes, based on 
a fixed number of 750 gradient descent iterations for comparability.

Importantly, the runtime of EvoMap is comparable to that of static MDS applied independently 
to each time slice. As expected, static MDS tends to be faster, since it does not compute the temporal 
cost component. However, the difference is often small—and in some settings, EvoMap can actually 
be faster. In particular, for datasets with relatively few objects but many time points, EvoMap 
shows runtime advantages. This is because it solves a single optimization problem over the full sequence, 
rather than optimizing each configuration independently at each time step.

Finally, we note that the application presented in \cite{Matthe+Ringel+Skiera:2023} successfully scaled the method 
to roughly 1{,}000 firms over 20 years of data, which further demonstrates the package’s scalability.

%% -- Summary/conclusions/discussion -------------------------------------------

\section{Summary and discussion} \label{sec:summary}

The \pkg{evomap} package, showcased in this paper, equips users with a comprehensive toolbox for dynamic mapping,
enabling them to focus on their specific application rather than the method's implementation. The package supports
the full mapping workflow—from data preparation and model fitting to visualization and evaluation—and is designed
to be flexible across a wide range of potential use cases.

Despite its broad functionality, the package also comes with limitations—many of which reflect constraints
introduced by the currently implemented static mapping methods (MDS, Sammon mapping, and t-SNE).
For example, these methods require fully observed data at each time point. While \pkg{evomap} can accommodate
unbalanced panels (see Section~\ref{sec:unbalanced_data}), it does not support incomplete data in the conventional sense.
Still, users may address this issue in several ways: If only some observations are missing within a time point,
interpolation based on neighboring periods may be feasible. If entire variables are missing from the raw input
tabular data, one could exclude those variables altogether—though they should then be dropped across all periods
to ensure comparability. If data are missing only for a subset of objects in certain periods, treating the
dataset as unbalanced is another option. Finally, users may substitute the static mapping method with one that
handles missing data directly. 

These considerations point to several directions for future development. As outlined in Section~\ref{sec:package},
\pkg{evomap}'s modular architecture facilitates the integration of additional mapping methods. For instance, the 
currently implemented methods do not allow for asymmetric relationships, nor do they generate output other than Cartesian planes.
As static mapping methods exist that can accommodate these requirements, like asymmetric or
spherical MDS, adding them to the evomap package could further extend EvoMap's possible areas of application. Further, 
combining EvoMap with external preference analysis or property fitting could extend the resultant maps' interpretability.

While we focused this presentation on temporal sequences, the underlying logic of dynamic mapping is more general and 
could be applied to any ordered sequence of relational data that may benefit from alignment. This may include spatial 
sequences, such as data collected across neighboring locations or along a geographic path, as well as behavioral sequences,
such as measures of perceptions evolving along a customer journey. Dynamic mapping could also be used when time is measured on
the object level—such as by user age, developmental stages, or product versions, which would allow researchers to visualize
systematic change without requiring synchronized observation periods.

To conclude, we hope that the \pkg{evomap} package can aid users in applying the EvoMap framework, and that it can 
serve as a foundation for manifold applications of dynamic mapping in different domains.

\section*{Acknowledgments}

This research was supported by the German Research Foundation, Deutsche Forschungsgemeinschaft (grant number SK 66/9-1). 

The author would like to thank Alexsey Pechnikov for invaluable support in optimizing the code for this project. 

\bibliography{jss5549}

\end{document}